\documentclass[journal,twoside,print]{ieeecolor}

\usepackage{generic}
\usepackage{cite}
\usepackage{amsmath,amssymb,amsfonts}
\usepackage{algorithm,algpseudocode}
\usepackage{graphicx}
\usepackage{textcomp}

\usepackage{array}
\usepackage{stfloats}
\usepackage{url}
\usepackage{verbatim}
\usepackage{soul,color}
\usepackage{textcomp}
\usepackage{titlesec}
\usepackage{indentfirst}
\usepackage{subfigure}
\usepackage{multirow}
\usepackage{hhline}
\usepackage{graphicx}
\usepackage{subcaption}
\usepackage{booktabs}  
\usepackage{multirow}  
\usepackage{array}     

\makeatletter
\let\NAT@parse\undefined
\let\labelindent\relax 
\makeatother
\usepackage[shortlabels]{enumitem}
\usepackage[colorlinks,citecolor=green,urlcolor=blue,bookmarks=false,hypertexnames=true]{hyperref}

\def\BibTeX{{\rm B\kern-.05em{\sc i\kern-.025em b}\kern-.08em
    T\kern-.1667em\lower.7ex\hbox{E}\kern-.125emX}}
\begin{document}

\title{SpiRadar: Radar-Based Non-Contact Spirometry via Sparse Polynomial Framework}
\author{Yonathan Eder, Yhonatan Kvich, Naama Golan, Adi Wagerhoff, Safit Levy,  Lior Barbash, Ron Berant, Patrick Stafler, and Yonina C. Eldar \vspace{-1.0cm}
\thanks{
This research was supported by the European Research Council (ERC) under the European Union’s Horizon 2020 research and innovation program (grant No. 101000967), by the Israel Science Foundation (grant No. 536/22), by The Manya Igel Centre for Biomedical Engineering and Signal Processing, and by Dr. Gilbert S. Omenn and Martha A. Darling WISH Fund for Clinical Breakthroughs through Scientific Collaboration \textit{(Corresponding author: Yonathan Eder.)}\\
Yonathan Eder, Yhonatan Kvich, Adi Wagerhoff, Safit Levy, and Yonina C. Eldar are with the Faculty of Mathematics and Computer Science, Weizmann Institute of Science, Rehovot, Israel (e-mail: yoni.eder@weizmann.ac.il, yonatan.kvich@weizmann.ac.il, adi.wegerhoff@weizmann.ac.il, safit.levy@weizmann.ac.il, yonina.eldar@weizmann.ac.il).\\
Naama Golan, Lior Barbash, Ron Berant, and Patrick Stafler are with the Pulmonary Institute, Schneider Children's Medical Center of Israel, Petah Tikva. Naama Golan, and Patrick Stafler are also with the Gray Faculty of Medical \& Health Sciences, Tel Aviv University, Tel Aviv, Israel (e-mail: naama.be.nbe@gmail.com, liorbarbash@gmail.com, ron.berant@gmail.com, pstafler@hotmail.com).}   }

\maketitle

\begin{abstract}
Chronic respiratory diseases affect hundreds of millions of people worldwide, with spirometry serving as the gold standard for pulmonary function assessment. However, conventional spirometry's reliance on mouthpieces and nose clips creates discomfort and technical challenges that can compromise test quality, particularly in pediatric populations where cooperation difficulties are amplified. This study presents SpiRadar, a comprehensive radar-based framework for non-contact spirometry that eliminates physical contact requirements while enabling accurate curve reconstruction, clinical parameter estimation, and bronchodilator response (BDR) assessment. Our key contributions include: (1) a methodological framework integrating physiologically-motivated preprocessing for robust signal extraction, a feature-dependent polynomial transformation linking radar-measured thoracic displacement to spirometric volume curves, and sparse optimization enabling generalization to unseen subjects without individual calibration; (2) a rigorous validation on a pediatric cohort of 39 subjects (ages 6–18 years) undergoing 58 spirometry trials, including healthy children and asthma patients tested pre- and post-bronchodilator, using subject-level leave-one-out cross-validation; (3) clinical-grade performance outperforming alternative non-contact methods, achieving accurate curve reconstruction (mean RMSE: 0.23±0.12 L) and strong correlations ($r>0.84$) for key spirometric parameters. BDR classification achieved 89.5\% accuracy with balanced sensitivity (90.9\%) and specificity (87.5\%). These results establish clinical feasibility for our non-contact spirometry approach, with particular promise for pediatric asthma monitoring and remote patient care.
\end{abstract}


\section{Introduction}
\label{sec:introduction}

\IEEEPARstart{R}{espiratory} diseases, including asthma and chronic obstructive pulmonary disease (COPD), constitute a substantial global health burden with profound clinical and economic impacts, costing the United States healthcare system tens of billions of dollars annually \cite{ferkol2014global, nurmagambetov2018economic, guarascio2013clinical}. Pulmonary function testing (PFT), particularly spirometry, serves as the gold standard for diagnosing and monitoring respiratory diseases, providing objective measurements of lung capacity and airflow limitation \cite{miller2005standardisation,graham2019standardization}. Among pediatric populations, asthma remains the most prevalent chronic respiratory condition, necessitating regular monitoring to optimize treatment and prevent exacerbations \cite{gaillard2021european}. 

However, conventional spirometry presents unique challenges that are heightened in pediatric populations. Children often struggle with the technical demands of forced expiratory maneuvers, including proper mouthpiece seal maintenance, nose clip tolerance, and the sustained cooperation required for reproducible measurements \cite{seed2012children}. These challenges are particularly pronounced in younger children, with current guidelines typically limiting reliable spirometry to children aged 5 years and above \cite{gaillard2021european, graham2019standardization}. These difficulties can lead to suboptimal test quality, multiple failed attempts, and patient distress, ultimately limiting the utility of spirometry in young patients \cite{gaillard2021european}. Furthermore, the COVID-19 pandemic has highlighted the need for alternative testing approaches that minimize close contact between patients and healthcare providers, while home-based monitoring has emerged as an important consideration for managing chronic respiratory conditions \cite{khor2025assessment}.

Recent advances in radar technology have opened new possibilities for non-contact physiological monitoring. Millimeter-wave (mmWave) frequency-modulated continuous wave (FMCW) radar systems can detect subtle chest wall movements associated with respiration and heartbeat, offering several advantages over traditional contact-based methods \cite{fioranelli2019radar, liebetruth2024systematic}. Recent work has demonstrated robust multi-person vital signs monitoring \cite{eder2023sparsity, eder2025robust} with sparsity-based signal processing \cite{eldar2012compressed,eldar2015sampling}, achieving accurate heart rate and respiratory rate extraction even in noisy, cluttered environments. Building upon these advances in radar-based physiological monitoring, the application of such techniques to more complex respiratory assessments, such as spirometry, represents a natural progression toward comprehensive pulmonary function evaluation.

Despite recent advances, existing approaches linking non-contact signals with spirometry measurements remain limited. 
Massagram et al. \cite{massagram2013tidal} demonstrated a linear relationship between radar-extracted chest displacement and spirometry volume, though their work focused on normal breathing rather than forced spirometry maneuvers and required prior knowledge of the spirometry reference's standard deviation, limiting clinical applicability. Liu et al. \cite{liu2017noncontact} established a fifth-order polynomial relationship between webcam-recorded shoulder displacement and spirometry volume; however, their method required subject-specific calibration using six breathing maneuvers, preventing generalization to unseen subjects. Wang et al. \cite{wang2023millimeter} employed linear regression to directly estimate spirometric parameters from radar-derived and anthropometric features, bypassing curve reconstruction entirely. While computationally efficient, this approach sacrifices the diagnostic value of complete spirometry waveforms.

A critical limitation across these studies is their insufficient emphasis on the morphological characteristics of volume-time and flow-volume curves, which contain crucial diagnostic value \cite{stanojevic2022ers}. 
Clinicians depend on these visual patterns for quality control of maneuver performance and for diagnostic interpretation, including early disease detection, severity assessment, and distinguishing obstructive from restrictive conditions. Moreover, bronchodilator response (BDR) assessment is a critical measure to estimate the reversibility of airway obstruction for patients who may have asthma or COPD \cite{waalkens1993assessment,graham2019standardization}. However, it remains unaddressed in prior non-contact spirometry research. These gaps highlight the need for a comprehensive, generalizable framework capable of accurate estimation of spirometry curves, clinical parameters, and therapeutic response assessment.

In this work, we present SpiRadar, a robust framework for non-contact spirometry and BDR assessment using mmWave FMCW radar. SpiRadar employs a hybrid learning-based approach that leverages sparse optimization via a feature-dependent polynomial model to learn generalizable coefficient mappings from training data. Our framework achieves robust generalization with pediatric training data (39 subjects, 58 trials), including healthy children and asthma patients tested pre- and post-bronchodilator (BD), through subject-level leave-one-out cross-validation (LOOCV). Once trained on a diverse cohort, the learned coefficient vector enables direct application to new subjects without individual calibration by a spirometer. This calibration-free design is particularly advantageous for clinical deployment, where per-subject training would be impractical. Our approach addresses existing methodological limitations through the following primary contributions:

\begin{itemize}
\item A methodological framework that integrates: (a) physiologically-motivated preprocessing with logarithmic interpolation correction to address involuntary movement artifacts unique to forced spirometry maneuvers; (b) a feature-dependent polynomial model linking radar-measured thoracic displacement to spirometric volume, incorporating anthropometric and radar-derived features; and (c) a sparse optimization framework enabling generalization to unseen subjects without individual calibration.

\item Comprehensive validation in a pediatric cohort of 39 subjects (ages 6–18 years) undergoing 58 spirometry trials, including healthy children and asthma patients tested pre-BD and post-BD, using rigorous subject-level LOOCV.

\item Superior performance compared to alternative non-contact methods: curve reconstruction (mean RMSE: 0.23±0.12 L), key spirometric parameters: forced vital capacity (FVC, $r=0.902$) and forced expiratory volume in 1 second (FEV$_1$, $r=0.846$), as well as BDR classification (accuracy: 89.5\%, sensitivity: 90.9\%, specificity: 87.5\%).
\end{itemize}

The remainder of this paper is organized as follows. Section \ref{sec:signal_models} presents the FMCW radar signal model and the proposed polynomial radar-spirometry relationship. Section \ref{sec:SpiRadar} details the SpiRadar methodology, including preprocessing and sparse optimization framework. Section \ref{sec:experimental_setup} describes the experimental setup and validation design. Section \ref{sec:results} presents comprehensive results for curve reconstruction, parameter estimation, BDR assessment, and performance sensitivity analysis. Section \ref{sec:discussion} discusses the implications and limitations, and Section \ref{sec:conclusion} concludes the paper.

\section{Signal Model and Problem Formulation}
\label{sec:signal_models}
This section presents the mathematical signal model of FMCW radar and the relation to spirometry volume curves. Then, it formulates the underlying estimation problem.

\subsection{FMCW Radar Model}
\label{sec:Radar_Model}
Consider an mmWave FMCW radar system transmitting ${L^{(v)}}$ frames of chirp signals \cite{iovescu2020fundamentals} at frame rate $f_s$ toward the chest wall of a subject performing spirometry. The measurement occurs in a clinical environment containing clutter, where the thoracic peak point is located at distance-angle coordinates $\{d_0,\theta_0\}$ relative to the radar antennas. Within its field of view (FOV), the radar detects $U\geq1$ objects distributed across various range-angle positions.

For each frame, reflected echo signals are mixed with transmitted signals and digitized via analog-to-digital conversion (ADC), producing length-$N$ discrete baseband (\textit{beat}) signals \cite{alizadeh2019remote, turppa2020vital}. The resulting amplitudes are impacted by the radar cross-section (RCS) of the detected objects. We employ a single-input multiple-output (SIMO) uniform linear array (ULA) configuration with one transmitter and $K>1$ receivers positioned at ${r_k} \triangleq (k - 1)\lambda_{\text{max}}/2$ for $k = 1, \ldots, K$, where $\lambda_{\text{max}}$ denotes the chirp's maximal wavelength. We note that angular resolution and signal-to-noise ratio (SNR) can be enhanced using a multiple-input multiple-output (MIMO)  configuration, which can be equivalently represented as a SIMO model through orthogonal multiplexing techniques, e.g., time-division multiplexing \cite{eder2025robust,li2021signal}.

Building upon our previous formulations \cite{eder2024localization,eder2025ralu,eder2025robust}, we extract thoracic displacement for spirometry estimation through the following bilinear model:
\begin{equation}
\label{Y=AXB+W}
{\bf{Y}}_l = {\bf{A}}{\bf{X}}_l{{\bf{B}}} + {\bf{W}}_l,\quad l = 1,\ldots,{L^{(v)}},
\end{equation}
where ${\bf{Y}}_l\in \mathbb{C}^{N\times{K}}$ represents the radar data matrix for frame $l$, and ${\bf{A}}\in \mathbb{C}^{N\times M}$ is the range-related Vandermonde dictionary with entries ${\bf{A}}(n,m) \triangleq \exp(j2\pi f_m nT_f)$. Here, $T_f$ denotes the fast-time sampling interval, and $\{f_m\}_{m=1}^M$ are $M \geq U$ distinct beat frequencies corresponding to radial distances $\{d_m\}_{m=1}^M$ through
\begin{equation}
    \label{fm_dm}
f_m \triangleq \frac{2S}{c}d_m,\quad m=1,\ldots,M,
\end{equation}
where $c$ is the speed of light and $S \triangleq B/T_c$ represents the chirp slope, with bandwidth $B$ and duration $T_c$. The radial distance to the subject's thorax $d_0 \in \{d_m\}_{m=1}^M$.

The angle-related dictionary ${\bf{B}} \in \mathbb{C}^{P \times K}$ has entries ${\bf{B}}(p,k) \triangleq \exp(j\phi_p[k])$, where phase shifts $\{\phi_p[k]\}_{p=1}^P$ arise from the ULA geometry for $P \geq K$ azimuth angles $\{\theta_p\}_{p=1}^P$:
\begin{equation}
\label{antenna_phase}
\phi_p[k] \triangleq \frac{2\pi}{\lambda_{\text{max}}}r_k\sin\theta_p,\quad k = 1, \ldots ,K.
\end{equation}
The azimuth angle to the subject's thorax $\theta_0 \in \{\theta_p\}_{p=1}^P$. The noise matrix ${{\bf{W}}_l} \in \mathbb{C}^{N \times K}$ contains i.i.d. complex Gaussian elements with zero mean and variance $\sigma^2$.

The unknown matrix ${\bf{X}}_l \in \mathbb{C}^{M \times P}$ represents reflected amplitudes from up to $MP$ range-angle pairs, including the target thoracic region. Its entries are
\begin{equation}
    \label{complex_amplitudes}
    {\bf{X}}_l(m,p) \triangleq x_{m,p}\exp(j\psi_{m,p}[l]),\quad l = 1, \ldots ,{L^{(v)}},
\end{equation}
where the phase term
\begin{equation}
    \label{psi_l_human}
\psi_{m,p}[l] \triangleq \frac{4\pi}{\lambda_{\text{max}}}(d_m + v_{m,p}[l]),\quad l = 1, \ldots ,{L^{(v)}},
\end{equation}
incorporates the vibration function $v_{m,p}[l]$ which captures millimeter-scale thoracic motion $v[l]$ for appropriate $\{m,p\}$ location index. While our previous work successfully modeled periodic tidal breathing using sinusoidal decomposition \cite{eder2023sparsity}, the non-periodic forced expiratory maneuvers in spirometry necessitate a different approach, which we introduce in Section~\ref{sec:spirometry-radar_model}. 

\subsection{Spirometry Measurement Formulation}
\label{spirometry_prelim}
A typical contact spirometry test requires the subject to inhale maximally and then exhale forcefully for as long as possible using a dedicated mouthpiece and nose clip. The spirometer measures the air flow of the subject, which is then translated to discrete flow and volume curves, sampled at frequency $f_s$. These curves reveal critical information about disease type and severity and are used to calculate clinical parameters that quantitatively assess pulmonary function.

Let the discrete spirometry volume curve be denoted by $s[l]$, $l=1,...,L^{(s)}$ and the sub-segment of expiration by $s_{ex}[l]$, $l=1,...,L^{(s)}_{ex}$, where $\{s_{ex}[l]\}_{l=1}^{L^{(s)}_{ex}}\in \{s[l]\}_{l=1}^{L^{(s)}}$ and  $L^{(s)}_{ex} \leq L^{(s)}$. The expiration volume signal $s_{ex}[l]$ represents a monotonically increasing function starting from zero of accumulated liters of expelled air. Four key clinical parameters following ATS/ERS guidelines \cite{miller2005standardisation} can be computed from $s_{ex}[l]$:

\begin{equation}
\label{spir_params_calc}
\begin{aligned}
\text{FVC} &= \max_l\{s_{ex}[l]\} &&\text{[L]} \\
\text{FEV}_1 &= s_{ex}[f_s] &&\text{[L]} \\
\text{FEV}_1/\text{FVC} &= \frac{\text{FEV}_1}{\text{FVC}} \times 100 &&\text{[\%]} \\
\text{PEF} &= \max_l\{\dot{s}_{ex}[l]\} &&\text{[L/s]}
\end{aligned}
\end{equation}
where $\dot{s}_{ex}[l] = f_s \cdot (s_{ex}[l+1] - s_{ex}[l])$ represents instantaneous flow computed via finite differences and [L] stands for liter units. These parameters represent: Forced Vital Capacity (total exhaled volume), Forced Expiratory Volume in 1 second (volume after 1s of forced maneuver), their ratio in percentage, and Peak Expiratory Flow (maximum flow rate).

The clinical necessity for complete curve visualization, combined with the ability to derive all parameters from the reconstructed curves, motivates our approach to estimate the entire spirometry curve from radar rather than predicting only individual parameters.

\subsection{Proposed Spirometry-Radar Model}
\label{sec:spirometry-radar_model}
 \subsubsection{Feature-Dependent Polynomial Relationship}
Consider a radar system that records thoracic displacement of a subject during a forced spirometry maneuver. The discrete thoracic movement during expiration is denoted by $v_{ex}[l]$, $l=1,\ldots,L^{(v)}_{ex}$, where $\{v_{ex}[l]\}_{l=1}^{L^{(v)}_{ex}} \in \{v[l]\}_{l=1}^{L^{(v)}}$ (\ref{psi_l_human}) and ${L^{(v)}_{ex}} \leq {L^{(v)}}$. The radar-based expiration segment $v_{ex}[l]$ is considered as a monotonically increasing function starting from zero, describing the degree to which the torso moves away from the radar during expiration.

We model the relationship between thoracic displacement and spirometry volume during expiration using a feature-dependent polynomial model. The model consists of an $I$-th order polynomial transformation, where $I$ is selected to balance model expressiveness and generalization capability (optimal $I=3$ determined through sensitivity analysis in Section~\ref{poly_order_analysis}):

\begin{equation}
s_{ex}[l] = \sum_{i=1}^{I} \beta_i (v_{ex}[l])^i, \quad l = 1, \ldots, L_{ex}, \label{eq:poly_model}
\end{equation}

where $L_{ex} = \min(L^{(s)}_{ex}, L^{(v)}_{ex})$, and the polynomial coefficients $\{\beta_i\}_{i=1}^{I}$ depend linearly on subject features:

\begin{equation}
\beta_i = \sum_{q=1}^{Q} c_{i,q}\alpha_q, \quad i = 1, \ldots, I, \label{eq:beta_features}
\end{equation}

where $\{\alpha_q\}_{q=1}^{Q}$ represents computed anthropometric and radar-derived features, while $\{c_{i,q}\}$ are the unknown coefficients to be learned.

The polynomial model in (\ref{eq:poly_model})--(\ref{eq:beta_features}) can be expressed in matrix form as

\begin{equation}
\mathbf{s} = \mathbf{V}\boldsymbol{\beta} = \mathbf{V}\mathbf{C}\boldsymbol{\alpha}, \label{eq:matrix_form}
\end{equation}

where $\mathbf{s} \in \mathbb{R}^{L_{ex} \times 1}$: $\mathbf{s} \triangleq [s_{ex}[1], \ldots, s_{ex}[L_{ex}]]^T$, the coefficient vector $\boldsymbol{\beta} \in \mathbb{R}^{I \times 1}$: $\boldsymbol{\beta} \triangleq [\beta_1, \ldots, \beta_I]^T$ and $\mathbf{V} \in \mathbb{R}^{L_{ex} \times I}$, $L_{ex} > I$ is a Vandermonde matrix of powers, whose entries are given by $\mathbf{V}(l,i) \triangleq (v_{ex}[l])^i$. Notably, by denoting the radar-based expiration vector $\mathbf{v} \in \mathbb{R}^{L_{ex} \times 1}$: $\mathbf{v} =[v_{ex}[1], \ldots, v_{ex}[L_{ex}]]^T$, we have that $\mathbf{V}$ is assembled by taking $I$ powers for $\mathbf{v}$, that is, $\mathbf{V} =[\mathbf{v}^1, \ldots, \mathbf{v}^I]$. $\mathbf{C} \in \mathbb{R}^{I \times Q}$ is the matrix of all unknown coefficients, whose entries are given by $\mathbf{C}(i,q) \triangleq c_{i,q}$, and $\boldsymbol{\alpha} \in \mathbb{R}^{Q \times 1}$: $\boldsymbol{\alpha} \triangleq [\alpha_1, \ldots, \alpha_Q]^T$ is a feature vector.

To facilitate the transformation model, we linearize (\ref{eq:matrix_form}) using the property for bilinear matrix multiplication that $\text{vec}(\mathbf{AXB}) = (\mathbf{B}^T \otimes \mathbf{A})\text{vec}(\mathbf{X})$, where $\otimes$ denotes the Kronecker product~\cite{petersen2008matrix}. By applying it on (\ref{eq:matrix_form}) we get

\begin{equation}
\mathbf{s} = \mathbf{D}\mathbf{c}, \label{eq:linear_form}
\end{equation}

where $\mathbf{D} \in \mathbb{R}^{L_{ex} \times IQ}$: $\mathbf{D} \triangleq \boldsymbol{\alpha}^T \otimes \mathbf{V}$, $L_{ex} > IQ$, and $\mathbf{c} \in \mathbb{R}^{IQ \times 1}$: $\mathbf{c} \triangleq \text{vec}(\mathbf{C}) = [c_{1,1}, \ldots, c_{I,1}, \ldots, c_{I,Q}]^T$.

\subsubsection{Concatenated Learning Model}

To learn a generalizable mapping from radar to spirometry, we use $J$ measurement triplets $\{\mathbf{s}^{(j)}, \mathbf{v}^{(j)}, \boldsymbol{\alpha}^{(j)}\}_{j=1}^{J}$ from multiple subjects. For each $j$, a design matrix $\mathbf{D}^{(j)}$ is constructed from $\mathbf{v}^{(j)}$ and $\boldsymbol{\alpha}^{(j)}$, similarly to (\ref{eq:linear_form}).

Concatenating all measurements yields the global learning model:

\begin{equation}
\tilde{\mathbf{s}} = \tilde{\mathbf{D}}\mathbf{c} \label{eq:concat_model}
\end{equation}

where $\tilde{\mathbf{s}} \in \mathbb{R}^{\tilde{L} \times 1}$: $\tilde{\mathbf{s}} \triangleq [{\mathbf{s}^{(1)}}^T, \ldots, {\mathbf{s}^{(J)}}^T]^T$, $\tilde{\mathbf{D}} \in \mathbb{R}^{\tilde{L} \times IQ}$: $\tilde{\mathbf{D}} \triangleq [{\mathbf{D}^{(1)}}^T, \ldots, {\mathbf{D}^{(J)}}^T]^T$, and $\tilde{L} = \sum_{j=1}^{J} L_{ex}^{(j)}$, where $L_{ex}^{(j)}$ denotes the length of the curves for the $j$'th trial.

Learning ${\mathbf{c}}$ from this multi-subject concatenated data $\{\tilde{\bf{s}},\tilde{\bf{D}}\}$, enables spirometry estimation for unseen subjects: given a new pair of $\{\mathbf{v}, \boldsymbol{\alpha}\}$, we estimate the spirometry curve ${\mathbf{s}}$ given the assembled $\mathbf{D}$ (\ref{eq:linear_form}) and the learned $\bf{c}$ (\ref{eq:concat_model}). 

\subsection{Assumptions and Problem Formulation}
The experimental procedure includes a spirometry session using both a radar sensor and a spirometer device as a reference. The spirometer records airflow for constructing $\mathbf{s}$ (\ref{eq:matrix_form}) and the FMCW radar simultaneously transmits, receives, and processes $L$ frames of chirp signals for constructing $\{\mathbf{Y}_l\}_{l=1}^{L}$ (\ref{Y=AXB+W}). We make the following assumptions on the acquired signals:

\textbf{SIMO FMCW Radar Model Assumptions:}
\begin{enumerate}[{A}-1]  
\item \label{A-1} \textit{Subject stationarity}: The thoracic coordinates $\{m, p\}$ in $\{\mathbf{X}_l\}_{l=1}^{L}$ corresponding to the physical location $\{d_0, \theta_0\}$ remain fixed and joint across all $L$ frames.
\item \label{A-2} \textit{Sparse target distribution}: The number of objects in the FOV, $U$, satisfies $U \ll MP$, making $\{\mathbf{X}_l\}_{l=1}^{L}$ $U$-sparse matrices.
\end{enumerate}

\textbf{Spirometry-Radar Model Assumptions:}
\begin{enumerate}[{B}-1]  
\item \label{B-1} \textit{Monotonicity}: The radar displacement vector $\mathbf{v}$ is monotonically increasing function starting from zero.
\item \label{B-2} \textit{Coefficient sparsity}: The coefficient vector $\mathbf{c}$ is sparse.
\end{enumerate}

Assumptions \ref{A-1} and \ref{A-2} facilitate clear localization maps for robust thoracic displacement extraction~\cite{eder2023sparsity}. Assumption \ref{B-1} ensures numerical stability through distinct samples analogous to cumulative volume measurements, while \ref{B-2} reflects that although $\mathbf{D}$ (\ref{eq:linear_form}) significantly expands the feature space, only a subset meaningfully contributes to the transformation, suggesting that sparse regularization may mitigate overfitting.

Our objectives are:
\begin{enumerate}
\item Learn a transformation mapping $\mathbf{c}$ that relates radar-measured thoracic displacement and subject features to spirometry volume curves across the population, given training data $\{\tilde{\bf{s}},\tilde{\bf{D}}\}$ (\ref{eq:concat_model}).
\item Reconstruct spirometry curves $\hat{\mathbf{s}}$ for unseen subjects from their radar measurements $\mathbf{v}$ and features $\boldsymbol{\alpha}$.
\item Extract clinical parameters (FVC, FEV$_1$, FEV$_1$/FVC ratio, PEF) from reconstructed curves.
\item Assess BDR for asthma patients using the pediatric threshold of $\geq12\%$ FEV$_1$ post-BD improvement from pre-BD baseline~\cite{jat2013spirometry}.
\end{enumerate}

\section{SpiRadar Methodology}
\label{sec:SpiRadar}
\begin{figure*}[htbp!]  
\begin{center}
\hspace{-0.0cm}
\subfigure{\label{}\includegraphics[width=1.0\textwidth]{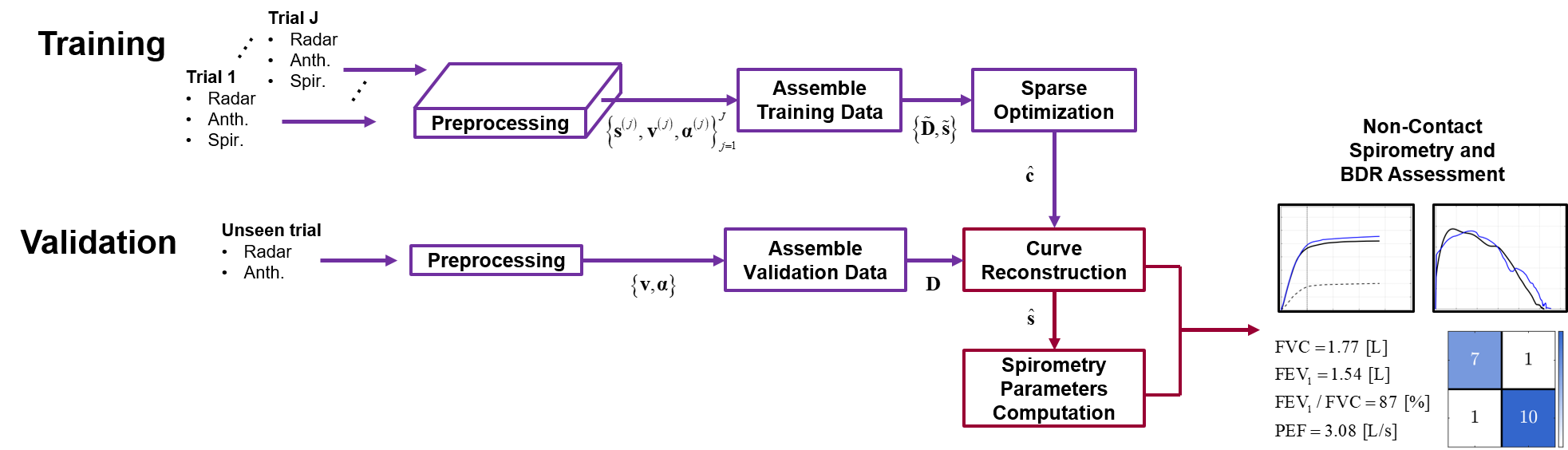}}
\end{center} 
\vspace{-0.5cm}
\caption{Block diagram of the proposed SpiRadar framework for radar-based non-contact spirometry and BDR assessment.}   
\label{fig:alg_block_diagram}
\vspace{-0.3cm}
\end{figure*}

This section details the implementation of the proposed SpiRadar framework, as depicted in Fig.~\ref{fig:alg_block_diagram}. To ensure rigorous validation and prevent data leakage, we employ subject-level LOOCV, where each fold holds out one subject (1-2 trials depending on health status) for testing while training on all remaining subjects' trials (elaborated in Section~\ref{sec:validation}). The following subsections describe the preprocessing and sparse estimation procedures applied within each LOOCV fold.

\subsection{Preprocessing}
\label{Pre_Proc}
The preprocessing stage transforms raw data into the required model inputs. For $J$ training trials, the pipeline extracts triplets $\{\mathbf{s}^{(j)}, \mathbf{v}^{(j)}, \boldsymbol{\alpha}^{(j)}\}_{j=1}^J$ containing reference spirometry, radar thoracic displacement, and subject features. For held-out test trials, only pairs $\{\mathbf{v}, \boldsymbol{\alpha}\}$ are extracted. The pipeline consists of three parallel processing paths detailed below and summarized in Algorithm~\ref{alg_preprocessing}.

\subsubsection{Radar Preprocessing}
\label{sec:radar_preprocessing}

\textbf{Data Assembly and Discretization.}
The raw radar channel data is processed to construct high-SNR input matrices $\{\mathbf{Y}_l\}_{l=1}^L$ conforming to the bilinear model in (\ref{Y=AXB+W}). To reduce noise variance by a factor of $G$, we coherently combine $G>1$ consecutive chirps per frame, exploiting the slowness of thoracic movement relative to frame period $T_s$ \cite{alizadeh2019remote, eder2023sparsity}. The range dictionary $\mathbf{A}$ is constructed on the Nyquist grid with frequencies 
\begin{equation}
    \label{fm_Nyquist_grid}
f_m  =\frac{{{f_{\textrm{ADC}}}}}{N} {i_m}{\rm{,}}\quad {i_m} = 0,\ldots,M - 1,
\end{equation}
where $f_{\textrm{ADC}}\triangleq1/T_f$ is determined by the ADC component and $M=N/2$ for single-sided spectrum. The angle dictionary $\mathbf{B}$ is discretized to cover a 180° FOV by
\begin{equation}
\label{theta_p_grid}
{\theta _p} =  - 90 + {i_p}{\Delta _\theta },\quad {i_p} = 0, \ldots ,P - 1,\quad P = \frac{{180}}{{{\Delta _\theta }}},
\end{equation}
where ${\Delta _\theta }$ denotes angle grid spacing.

\textbf{Localization and Displacement Extraction.}
To isolate thoracic motion from environmental clutter, we apply slow-time bandpass filtering on $\{{\mathbf{Y}}_l\}_{l=1}^L$ according to the vital band in \cite{eder2023sparsity,eder2025robust}. The filtered data $\{\bar{\mathbf{Y}}_l\}_{l=1}^L$ undergoes joint sparse recovery (JSR) exploiting assumptions \ref{A-1}, \ref{A-2}:
\begin{equation}
    \label{l_{2,1}_LS}
\hat{\mathbb{X}}= \mathop {\arg \min }\limits_{\mathbb{X} \in \mathbb{C}^{M\times{P}\times{L}}} \frac{1}{{2L}}\sum\limits_{l = 1}^L {\left\| {{{\bf{\bar{Y}}}_l} - {\bf{A}}{{\bf{X}}_l}{\bf{B}}} \right\|_F^2}  + \lambda {{\left\| \mathbb{X} \right\|}_{2,1}},
\end{equation}
where $\lambda\geq 0$ is the regularization parameter and ${\left\| \mathbb{X} \right\|_{2,1}}$ denotes the sum of $l_2$ norms over the frame dimension of the tensor $\mathbb{X} \in \mathbb{C}^{M\times{P}\times{L}}$ concatenating ${\left\{ {{{\bf{X}}_l}} \right\}_{l = 1}^L}$. This is solved using the RaLU-JSR localization algorithm proposed in \cite{eder2025robust}.

\begin{figure}[htbp!]  
\begin{center}
\subfigure{\includegraphics[width=0.35\textwidth]{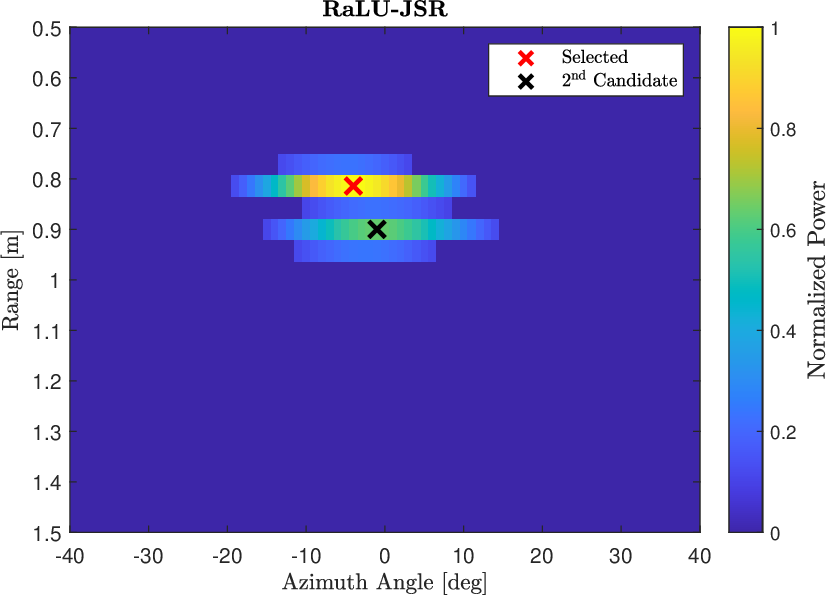}}\vspace{-0.2cm}
\end{center} 
\vspace{-0.2cm}
\caption{Normalized range-angle localization map of trial 4, showing multiple candidate locations. The selected optimal location is marked with a red {\color{red}{X}} sign, remaining candidates with black X.}   
\label{fig:loc_map}
\vspace{-0.2cm}
\end{figure}

From recovered $\hat{\mathbb{X}}$, we compute the frame-averaged power map $\bar{\mathbf{X}}(m,p) = (1/L)\sum_{l=1}^L |\hat{\mathbb{X}}(m,p,l)|^2$, normalize by the maximum within range-angle region of interest (ROI) of [0.5, 1.5] [m] and [-40, +40] [$^{\circ}$] to focus on the subject's thoracic area, and denoise by zeroing values below 30\% of the maximum. Since spirometry involves significant torso movement, the map may exhibit smearing across multiple range bins, producing multiple candidate locations (Fig.~\ref{fig:loc_map}). For each $(m,p)$ candidate, we compute the complex signal via beamforming $\hat{x}_{m,p}[l] = (1/NK)\mathbf{A}_m^H\mathbf{Y}_l\mathbf{B}_p^H$ and extract the vibration $\hat{v}_{m,p}[l] = \text{unwrap}(\angle(\hat{x}_{m,p}[l]))$, where $\text{unwrap}(\cdot)$ and $\angle(\cdot)$ denote phase unwrapping \cite{alizadeh2019remote} and four-quadrant arctangent, respectively.

\textbf{Expiration Segment Extraction.}
As demonstrated in Fig. \ref{fig:spirometry_radar_comp}(b), each $\hat{v}_{m,p}[l]$ candidate undergoes: (1) Savitzky-Golay smoothing (degree 2, window $f_s$) \cite{schafer2011savitzky}; (2) Back-Extrapolated Volume (BEV) point detection via peak-flow extrapolation following ATS/ERS guidelines \cite{graham2019standardization}. The BEV point serves as Time 0 for standardized measurements, accounting for initial hesitation during maneuvers. (3) Truncation from BEV to subsequent minimum within 7s to capture the complete expiratory pattern. The BEV truncation eliminates the initial plateau region for improving the monotonicity of the radar-to-spirometry transformation; (4) signal correction due to artifacts that arise from involuntary body movements and the involvement of multiple torso organs that do not correspond to volume changes. Motivated by physiological expiratory flow decay \cite{miller2005standardisation}, we employ a logarithmic fit with coefficients determined by preserving expiration endpoints and slope before the non-monotonic segments. (5) flip and shift to zero-starting monotonically increasing form.

The optimal location $(\hat{m}, \hat{p})$ is selected based on the longest expiration duration detected, used as an indicator of signal quality. Interestingly, the selected signal exhibits additional temporal characteristics, including distinct breathing patterns and high contrast between the maneuver and baseline
tidal breathing, as seen in Fig. \ref{fig:spirometry_radar_comp}(b). These characteristics could serve as additional selection criteria in future implementations. The final radar displacement vector is $\mathbf{v} \triangleq [v_{\hat{m},\hat{p}}[1], \ldots, v_{\hat{m},\hat{p}}[L_{ex}^{(v)}]]^T$.

\subsubsection{Feature Vector Assembly}
The feature vector $\boldsymbol{\alpha}$ (\ref{eq:matrix_form}) combines anthropometric measurements with radar-derived respiratory parameters. Anthropometrics include gender (one-hot encoded as Gender$_0$, Gender$_1$), age, height, weight, body mass index (BMI), and body surface area (BSA), calculated per spirometer standards \cite{SmartPFT_USB_2025}. Radar-derived parameters computed from processed $v_{ex}[l]$ include: FEV$_1^R = v_{ex}[f_s]$, ratio$^R = v_{ex}[f_s]/\max_l\{v_{ex}[l]\} \times 100\%$, and PEF$^R = \max_l\{f_s \cdot (v_{ex}[l+1] - v_{ex}[l])\}$. The FVC$^R$ parameter is intentionally excluded to prevent overfitting, following the analysis detailed in Section \ref{feature_analysis}. The final feature vector is then
\begin{equation}
\label{final_feature_vector}
\boldsymbol{\alpha} =
\begin{aligned}[t]
[&\text{Gender}_0, \text{Gender}_1, \text{Age}, \text{Height}, \text{Weight}, \\
 &\text{BMI}, \text{BSA}, \text{FEV}_1^{\text{R}}, \text{ratio}^{\text{R}}, \text{PEF}^{\text{R}}]^\top.
\end{aligned}
\end{equation}

\subsubsection{Spirometry Preprocessing (Training Only)}
\begin{figure*}[htbp!]  
\begin{center}
\subfigure{\includegraphics[width=0.7\textwidth]{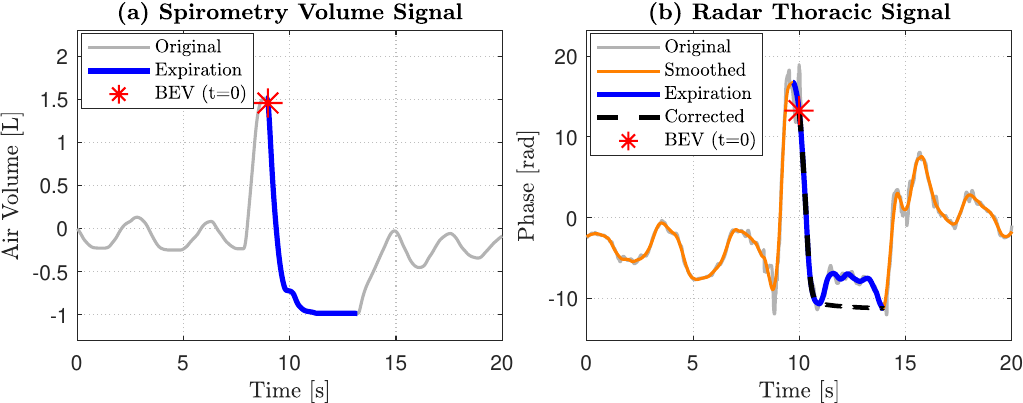}}\vspace{-0.2cm}
\end{center} 
\vspace{-0.2cm}
\caption{Preprocessing pipeline demonstration for trial 4 showing (left) reference spirometry volume signal and (right) corresponding radar thoracic vibration signal.}   
\label{fig:spirometry_radar_comp}
\vspace{-0.4cm}
\end{figure*}

For training subjects, reference spirometry undergoes the following preprocessing (Fig.~\ref{fig:spirometry_radar_comp}(a)): (1) BEV detection; (2) expiration truncation from BEV point to the subsequent minimum within 7s;  (3) flip and shift to zero-starting monotonically increasing form. The resulting vector is $\mathbf{s} \triangleq [s_{ex}[1], \ldots, s_{ex}[L_{ex}^{(s)}]]^T$. We note that for implementation, we ensure the radar and spirometry vectors are aligned in terms of length $L_{ex} = \min\{L_{ex}^{(s)}, L_{ex}^{(v)}\}$ and sampling rate of 125 [Hz].

Algorithm~\ref{alg_preprocessing} summarizes the complete pipeline.\vspace{-0.2cm}

\begin{algorithm}[h!]
\caption{Preprocessing Pipeline}
\label{alg_preprocessing}
\begin{algorithmic}
\State \textbf{Input:} Radar $\{\mathbf{Y}_l\}_{l=1}^L$, anthropometrics; \textbf{if training:} + spirometry $\{s[l]\}_{l=1}^{L^{(s)}}$

\State \textbf{// Radar Preprocessing}
\State \textbf{1.} Coherently combine $G$ chirps, construct $\mathbf{A}$ (\ref{fm_Nyquist_grid}), $\mathbf{B}$ (\ref{theta_p_grid})
\State \textbf{2.} Bandpass filter $\rightarrow \{\bar{\mathbf{Y}}_l\}$, solve (\ref{l_{2,1}_LS}) $\rightarrow \hat{\mathbb{X}}$
\State \textbf{3.} Compute map $\bar{\mathbf{X}}$, normalize/denoise (ROI, 30\% threshold)
\State \textbf{4.} \textbf{For each} $(m,p)$ \textbf{candidate:}
\State \quad \textbf{4.1.} Beamform: $\hat{x}_{m,p}[l]=(1/NK)\mathbf{A}_m^H\mathbf{Y}_l\mathbf{B}_p^H$
\State \quad \textbf{4.2.} Extract displacement: $\hat{v}_{m,p}[l] = \text{unwrap}(\angle(\hat{x}_{m,p}[l]))$
\State \quad \textbf{4.3.} Smooth, detect BEV, truncate, apply log correction, flip/shift
\State \textbf{5.} Select optimal $(\hat{m},\hat{p})$ by longest expiration $\rightarrow \mathbf{v}$

\State \textbf{// Feature Vector Assembly}
\State \textbf{6.} Compute BMI, BSA from anthropometrics
\State \textbf{7.} Compute FEV$_1^R$, ratio$^R$, PEF$^R$ from $v_{ex}[l]$
\State \textbf{8.} Assemble $\boldsymbol{\alpha}$ (\ref{final_feature_vector})

\If{training}
    \State \textbf{// Spirometry Preprocessing}
    \State \textbf{9.} Detect BEV, truncate, flip/shift $\rightarrow \mathbf{s}$
    \State \textbf{10.} Align $\mathbf{s}$ and $\mathbf{v}$ in length and sampling rate
    \State \textbf{Return:} $\{\mathbf{s}, \mathbf{v}, \boldsymbol{\alpha}\}$
\Else
    \State \textbf{Return:} $\{\mathbf{v}, \boldsymbol{\alpha}\}$
\EndIf
\end{algorithmic}
\end{algorithm}

\subsection{Proposed Sparse Estimation}
\label{c_est}
For each training session in the LOOCV framework, we concatenate the available training triplets $\left\{ {{{\bf{s}}^{\left( j \right)}},{{\bf{v}}^{\left( j \right)}},{{\boldsymbol{\alpha }}^{\left( j \right)}}} \right\}_{j=1}^J$ to follow the model $\tilde{\mathbf{s}} = \tilde{\mathbf{D}}\mathbf{c}$ (\ref{eq:concat_model}). Our objective is to learn the coefficient vector ${\bf{c}}$ that connects the concatenated design matrix ${\bf{\tilde{D}}}$ with the corresponding spirometry curves ${\bf{\tilde{s}}}$, such that ${\bf{c}}$ can generalize to relate any given design matrix ${\bf{D}}$ to its corresponding spirometry curve ${\bf{s}}$.

Assumption \ref{B-1} and the Kronecker structure of ${\bf{\tilde{D}}}$ provide distinct values that improve invertibility when using the standard Least-Squares (LS) approach \cite{ruppert1994multivariate}. However, to achieve better generalizability by avoiding overfitting and reducing the effect of outliers, we leverage the sparsity assumption of ${\bf{c}}$ in \ref{B-2}, which implies that many atoms in ${\bf{\tilde{D}}}$ do not contribute significantly to the radar-spirometry relationship in (\ref{eq:concat_model}).

Therefore, we propose to estimate ${\bf{c}}$ given ${\bf{\tilde{s}}}$ and ${\bf{\tilde{D}}}$ using the following $\ell_1$-norm LASSO formulation \cite{tibshirani1996regression}:
\begin{equation}
    \label{LASSO}
{\bf{\hat c}} = \mathop {\arg \min }\limits_{{\bf{c}} \in \mathbb{R}{^{IQ \times 1}}} \left\{ {\left\| {{\bf{\tilde s}} - {\bf{\tilde Dc}}} \right\|_2^2 + \gamma {{\left\| {\bf{c}} \right\|}_1}} \right\},
\end{equation}
where $\gamma\geq 0$ is the regularization parameter that controls the sparsity of ${\bf{c}}$ and correspondingly influences the goodness of fit. We solve this convex optimization problem using the Fast Iterative Soft-Thresholding Algorithm (FISTA) \cite{beck2009fast,palomar2010convex}, which has demonstrated effectiveness in our previous radar-based sparse recovery applications \cite{eder2023sparsity,eder2025robust}. Once the coefficient vector $\hat{\mathbf{c}}$ is estimated from the training data, it is applied to unseen test subjects to reconstruct spirometry curves using $\hat{\mathbf{s}} = \mathbf{D}\hat{\mathbf{c}}$ (\ref{eq:linear_form}), from which clinical parameters (FVC, FEV$_1$, FEV$_1$/FVC ratio, PEF) are subsequently computed (\ref{spir_params_calc}) for assessment of pulmonary function as well as BDR for asthma patients.

In practice, to ensure numerical stability and balanced feature contributions across the heterogeneous feature vector $\boldsymbol{\alpha}$, we apply column-wise min-max normalization \cite{jain2011min} to the design matrix $\tilde{\bf{D}}$ prior to optimization. For unseen test subjects, the training-derived normalization parameters are applied to the validation design matrix $\mathbf{D}$ to maintain consistency.


\section{Experimental Setup}
\label{sec:experimental_setup}
This section describes the experimental configurations used to evaluate SpiRadar's performance.

\subsection{Pediatric Dataset Description}
The study was conducted at the pulmonary clinic of Schneider Children's Medical Center of Israel / Rabin Medical Center (RMC). Institutional Review Board (IRB) approval was obtained from both institutes (RMC 0417-23 and WIS 2214-1) prior to subject recruitment. Informed consent was obtained from each accompanying parent before enrollment.

The study enrolled children aged 6–18 years, divided into two groups. The first group included healthy participants from the general population with no significant underlying disease, current respiratory symptoms, or fever. The second group comprised children with physician-diagnosed asthma who were regularly followed at the clinic and had spirometric evidence of an obstructive ventilatory defect at the time of attendance. Children with significant comorbidities were excluded. Anthropometric characteristics of the subjects are given in Table \ref{tab:anthropometric}.

\begin{table}[htbp]
\centering
\caption{Anthropometric Characteristics}
\label{tab:anthropometric}
\begin{tabular}{lccc}
\hline
Parameter & Healthy & Asthmatic & All \\
\hline
Gender (M/F) & 10/10 & 14/5 & 24/15 \\
Age (years) & 11.7 ± 3.7 & 11.9 ± 3.2 & 11.8 ± 3.4 \\
Height (cm) & 144.7 ± 15.3 & 149.7 ± 17.3 & 147.2 ± 16.3 \\
Weight (kg) & 39.6 ± 13.5 & 46.1 ± 19.3 & 42.7 ± 16.7 \\
\hline
Sample size & 20 & 19 & 39 \\
\hline
\end{tabular}
\\[1ex]
\footnotesize{Values presented as mean ± standard deviation. M = Male, F = Female.}
\end{table}

\subsection{Experimental Procedures}
\begin{figure}[htbp!]  
\begin{center}
\hspace{-0.3cm}
\subfigure{\label{}\includegraphics[width=0.40\textwidth]{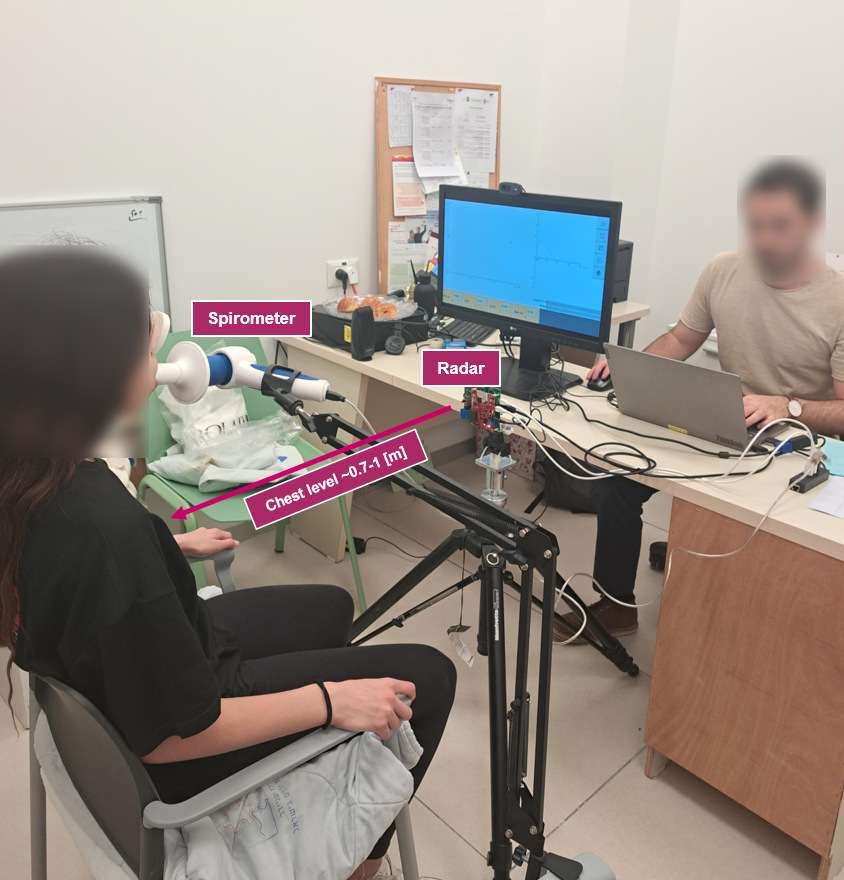}}
\end{center} 
\caption{Experimental procedure in a cluttered medical room. Each subject performed
spirometry testing in a seated position, which was simultaneously recorded by a spirometer device, serving as the GT reference, and a dedicated radar sensor positioned at the chest level, approximately 0.7–1 meters from the subject. }
\label{fig:exp_setup}
\end{figure}

As shown in Fig. \ref{fig:exp_setup}, all trials were conducted in a medical room containing typical clinical furniture, including a table, chairs, and a hospital bed. Each subject performed spirometry testing in a seated position, which was simultaneously recorded by a 125 [Hz] spirometer device (smart PFT USB - CE certified \cite{SmartPFT_USB_2025}) serving as the ground-truth (GT) reference, and a dedicated radar sensor (Texas Instruments IWR1443BOOST $76$ to $81$ [GHz] mmWave Sensor Evaluation Module (EVM) \cite{ti_iwr1443boost}) positioned at chest level approximately $0.7–1$ meters from the subject.

Each trial was supervised by a certified lung function technician who instructed subjects to breathe calmly before and after maneuvers and to minimize large body movements. Each session included several forced expiratory maneuvers performed in accordance with ATS/ERS criteria \cite{miller2005standardisation}. For the asthmatic group, radar and spirometry data were acquired both before and 15 minutes after BD administration. A total of 58 trials were performed and categorized into three groups: I Healthy (n=20), II Asthma Pre-BD (n=19), and III Asthma Post-BD (n=19), where asthmatic patients contributed paired pre- and post-BD measurements.

\subsection{Radar Configuration}
We employed the radar's horizontal MIMO ULA setup of $2$ transmitting antennas and $4$ receiving antennas \cite{li2021signal}. Since we consider trials of a single target in front of the radar, we operated the $2$ transmitters simultaneously to create a $1\times{K}$ SIMO array with increased SNR \cite{eder2025robust}. The main radar parameters, reflected in the signal model in (\ref{Y=AXB+W}) are summarized in Table \ref{table:FMCW_parameters}. We note that for bandwidth $B=ST_c\approx 4$ [GHz], the range resolution is $d_{\textrm{res}}=\frac{c}{2B}\approx 3.75$ [cm]. In addition, the radar parameters listed in Table \ref{table:FMCW_parameters} through (\ref{fm_dm}), (\ref{fm_Nyquist_grid}) and (\ref{theta_p_grid}) with angle grid spacing set as $\Delta_{\theta}=1$, enable coverage over a radial distance range of $d_{\rm{min}}=4.29$ [cm] to $d_{\rm{max}}=4.24$ [m] and an angular range of $\theta_{\rm{min}}=-90^\circ$ to $\theta_{\rm{max}}=+89^\circ$. 

\begin{table}[h!]
\caption{MIMO FMCW radar parameters}
\label{table:FMCW_parameters}
\centering
\begin{tabular}{ |l |c | c|  }
\hline
Parameter & Symbol & Value\\
\hline
Maximal chirp wavelength & $\lambda_{\textrm{max}}$   & $3.9$ [mm]\\
Chirp duration & $T_c$   & $57$ [$\mu \textrm{s}$]\\
ADC sampling rate & $f_{\textrm{ADC}}$   & $4$ [MHz]\\
Rate of frequency sweep & $S$   & $70$ [MHz/$\mu$s]\\
Frame duration & $T_s$   & $2$ [ms]\\
$\#$ of selected \textit{fast-time} samples  & $\bar{N}$   & $200$ \\
$\#$ of chirps per frame & $G$   & $40$ \\
$\#$ of transmitters & $J$   & $2$ \\
$\#$ of receivers & $K$   & $4$  \\
 \hline
\end{tabular}
\end{table}

\subsection{Validation Methodology}
\label{sec:validation}

To ensure rigorous evaluation and prevent data leakage, we implement subject-level LOOCV. In this approach, each subject's complete data (including all trials from that subject) is sequentially held out for testing while all remaining subjects form the training set. This subject-level partitioning is critical because asthma patients contribute paired pre- and post-BD trials; trial-level splitting would leak patient-specific characteristics between training and test sets, artificially inflating performance estimates, and is therefore avoided.

For each LOOCV fold, the pipeline proceeds as follows (Fig. \ref{fig:alg_block_diagram}): (1) preprocess raw data to extract $\{\mathbf{s}, \mathbf{v}, \boldsymbol{\alpha}\}$ for training subjects or $\{\mathbf{v}, \boldsymbol{\alpha}\}$ for held-out test subject's trials; (2) estimate coefficient vector $\mathbf{c}$ from concatenated training data; (3) reconstruct spirometry curve $\mathbf{s}$ for held-out trial using learned $\hat{\mathbf{c}}$; (4) compute clinical parameters and assess BDR when applicable.

\subsection{Evaluation Metrics and Comparison Methods}

We evaluate SpiRadar performance using multiple complementary metrics across three assessment categories:

\textbf{Curve Reconstruction:} Spirometry volume curves are assessed using Root Mean Square Error (RMSE) and coefficient of determination ($R^2$) to quantify reconstruction accuracy and goodness of fit.

\textbf{Parameter Estimation:} Clinical spirometry parameters (FVC, FEV$_1$, FEV$_1$/FVC ratio, PEF) are evaluated using RMSE, Mean Absolute Error (MAE), and Pearson correlation coefficient ($r$) to assess both absolute accuracy and linear relationship strength with reference values.

\textbf{BDR Classification:} For asthma patients, BDR assessment uses the standard pediatric threshold of $\geq12\%$ FEV$_1$ post-BD improvement from pre-BD baseline \cite{jat2013spirometry}. Classification performance is measured using accuracy, sensitivity, and specificity.

We compared SpiRadar against three baseline non-contact approaches: LinReg (linear regression for curve reconstruction \cite{massagram2013tidal}), PolyReg (polynomial regression for curve reconstruction \cite{liu2017noncontact}), and Direct (direct parameter estimation without curve reconstruction \cite{wang2023millimeter}). To ensure fair evaluation, all comparison methods received identical preprocessed data. Methods using a feature vector received SpiRadar's optimized $\boldsymbol{\alpha}$.

\section{Results}
\label{sec:results}
This section presents comprehensive evaluation results across curve reconstruction, parameter estimation, and BDR assessment, followed by sensitivity analyses of key algorithmic components.


\subsection{Curve Reconstruction}
\begin{figure}[htbp!]  
\begin{center}
\subfigure{\includegraphics[width=0.48\textwidth]{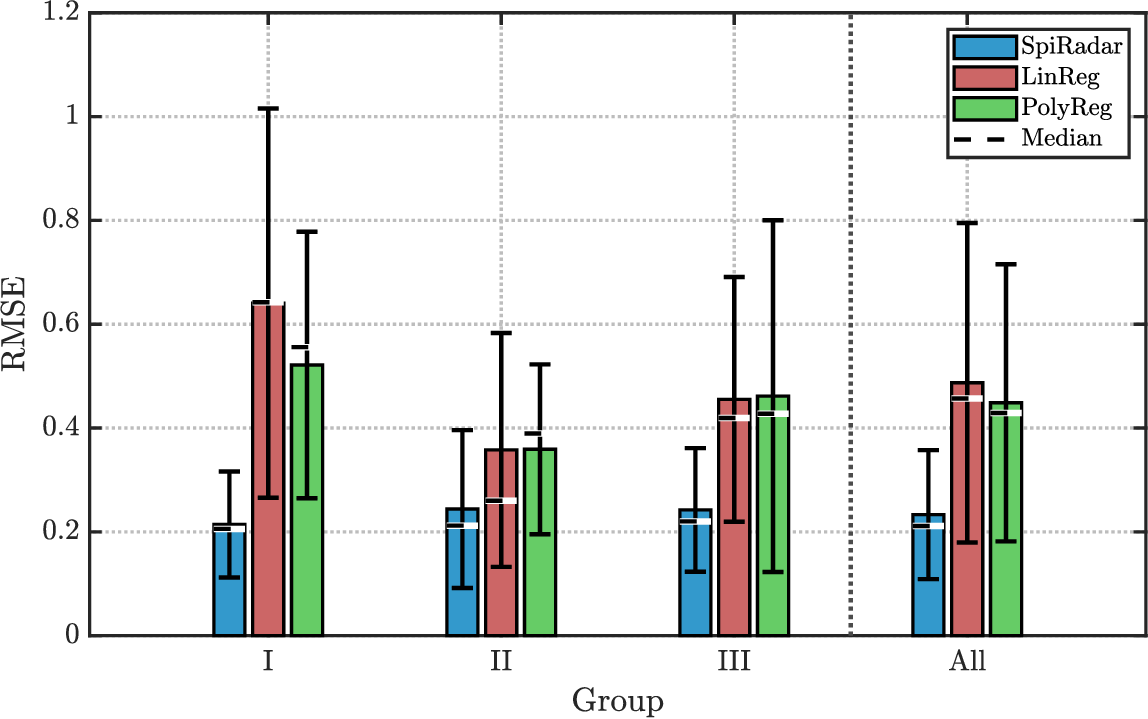}}\vspace{-0.2cm}
\end{center} 
\vspace{-0.2cm}
\caption{Curve fitting RMSE performance across clinical groups. Box plots show mean (box height), median (dashed line), and variability (whiskers) for each method. Groups: I-Healthy (n=20), II-Asthma Pre-BD (n=19), III-Asthma Post-BD (n=19), All-Combined (n=58). SpiRadar demonstrates the lowest RMSE ($<0.25$ L) across all groups.
\label{fig:rmse_by_group}
}
\label{fig:curve_rmse_group}
\vspace{-0.4cm}

\end{figure}

Fig. \ref{fig:curve_rmse_group} presents the overall curve fitting RMSE performance for all trials across the clinical groups. SpiRadar demonstrated superior and consistent performance with mean RMSE values below 0.25 [L] across all groups. The error bars show minimal variability for SpiRadar, with median values (dashed lines) closely aligned with means, indicating robust performance without significant outlier influence.

Table \ref{tab:curve_fit_overall} quantifies the overall performance in terms of both RMSE and R$^2$ metrics. SpiRadar achieved a mean RMSE of 0.23±0.12, representing a 53\% reduction compared to LinReg (0.49±0.31) and 49\% compared to PolyReg (0.45±0.27). The median RMSE for SpiRadar (0.21) was less than half that of the other methods. Critically, SpiRadar achieved positive R$^2$ values (mean: 0.68±0.41, median: 0.84), while LinReg and PolyReg showed negative mean R$^2$ values, suggesting limited effectiveness for spirometry curve modeling.

\begin{figure*}[htbp!]  
\begin{center}
\subfigure{\includegraphics[width=0.80\textwidth]{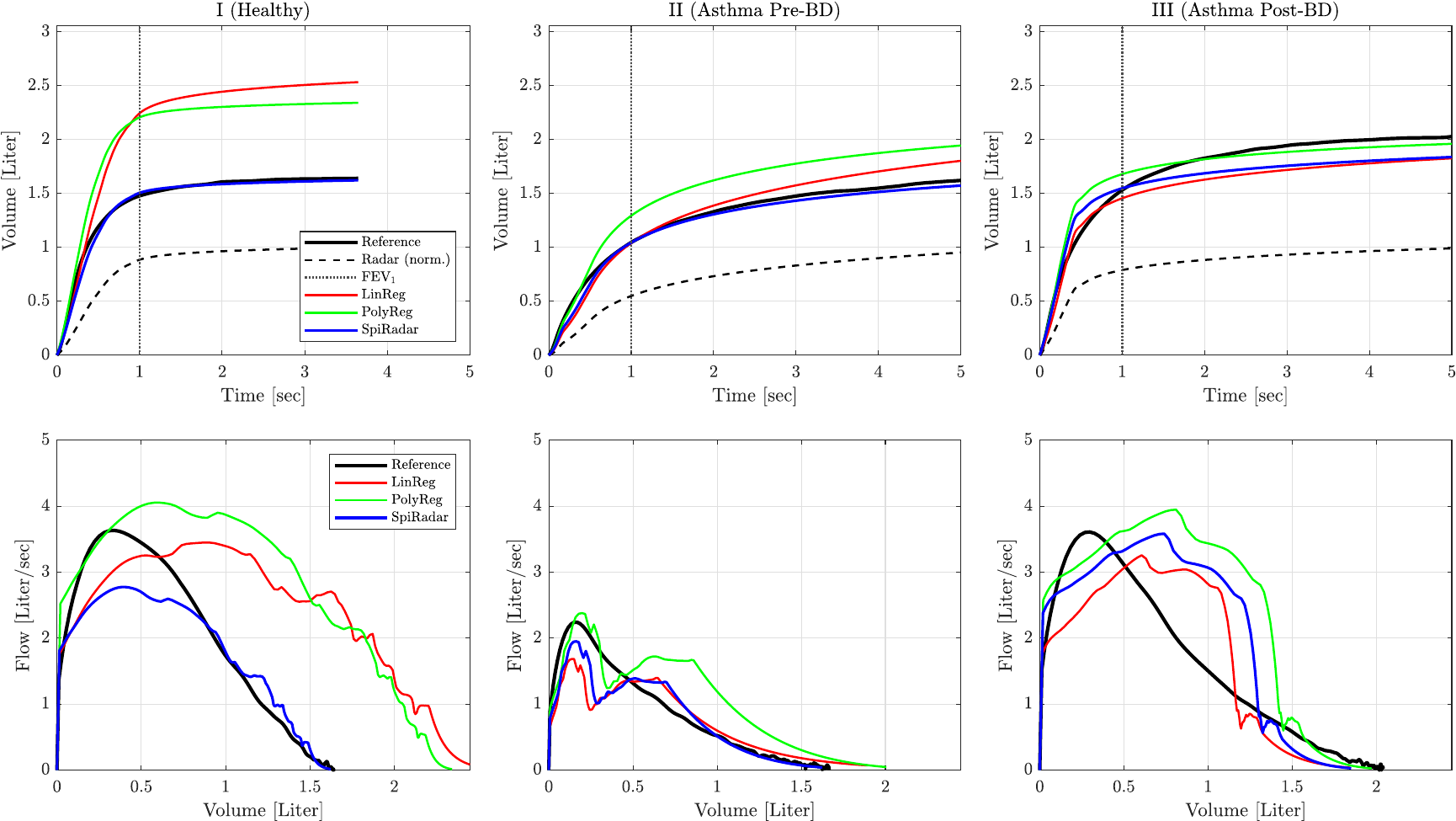}}\vspace{-0.2cm}
\end{center} 
\vspace{-0.2cm}
\caption{Representative reconstructed spirometry curves from each clinical group by the compared methods relative to the spirometry reference, showing volume-time (top) and flow-volume (bottom) curves. (I) Healthy (subject 31, trial 45), (II) Asthma Pre-BD (subject 39, trial 57), (III) Asthma Post-BD (subject 39, trial 58). 
}  
\label{fig:selected_curves}
\vspace{-0.4cm}
\end{figure*}

Fig. \ref{fig:selected_curves} illustrates representative spirometry curves from each clinical group. For the healthy subject (subject 31, trial 45), SpiRadar accurately tracked the reference curve and captured the true FEV$_1$ point at $1.5$ [L], while LinReg and PolyReg overestimated volumes throughout the expiration phase. The pre-BD asthma case (subject 39, trial 57) exhibited a characteristic obstructive pattern of slow expiration. While both LinReg and SpiRadar captured the reduced FEV$_1$ (1.07 [L]), only the latter accurately reproduced the complete volume-time trajectory. The corresponding post-BD trial (subject 39, trial 58) demonstrated a 46\% FEV$_1$ improvement (from 1.07 to 1.57 [L]), exceeding the 12\% pediatric threshold for significant response \cite{jat2013spirometry}. SpiRadar most accurately captured this improvement, while PolyReg's overestimations could lead to false-positive BD responses.

Flow-volume curve reconstruction presents inherent challenges due to flow being derived from volume data through differentiation, which amplifies noise and computational errors. Despite these difficulties, the flow-volume curves (bottom row) reveal important method differences. The characteristic concave pattern in pre-BD transformed to a more linear decline post-BD, with PEF improving from 2.0 to 3.5 [L/s] (75\% increase). SpiRadar captured both the flow pattern changes and PEF improvement most accurately, which is essential for distinguishing obstructive from restrictive patterns and assessing treatment response.



\begin{table}[!t]
\renewcommand{\arraystretch}{1.2}
\caption{Overall Curve Fitting Performance}
\label{tab:curve_fit_overall}
\centering
\setlength{\tabcolsep}{4pt}
\begin{tabular}{lcc}
\toprule
\multirow{2}{*}{Method} & \multicolumn{2}{c}{Mean$\pm$SD (Median)} \\
\cmidrule(lr){2-3}
& RMSE & $R^2$ \\
\midrule
LinReg & 0.49$\pm$0.31 (0.46) & $-$0.67$\pm$2.77 (0.30) \\
PolyReg & 0.45$\pm$0.27 (0.43) & $-$0.40$\pm$2.46 (0.31) \\
\textbf{SpiRadar} & \textbf{0.23$\pm$0.12 (0.21)} & \textbf{0.68$\pm$0.41 (0.84)} \\
\bottomrule
\end{tabular}
\vspace{-0.15in}
\end{table}

\subsection{Spirometry Parameters Estimation}
Table \ref{tab:performance_metrics} presents the performance metrics for spirometry parameter estimation across all 58 trials. Parameters were computed from estimated volume curves (LinReg, PolyReg, SpiRadar) and through direct parameter linear regression (Direct). As the best values are bolded, SpiRadar demonstrated superior performance across all spirometric parameters.

For FVC estimation, SpiRadar achieved the highest correlation ($r = 0.902$) with the lowest RMSE ($0.259$ [L]), representing $64\%$ and $59\%$ RMSE improvements compared to LinReg and PolyReg, respectively. While the Direct method showed strong correlation ($r = 0.881$), its error metrics remained higher than SpiRadar. Notably, PolyReg exhibited poor correlation ($r = 0.184$), indicating fundamental limitations in volume capacity estimation through polynomial regression.

FEV$_1$ estimation followed similar patterns. SpiRadar achieved $r = 0.846$ and RMSE = 0.283 [L], a $48\%$ and $41\%$ reduction compared to LinReg and PolyReg, respectively. Both regression methods showed correlations below $0.5$, highlighting their limited ability to accurately capture the critical one-second volume marker. The Direct method maintained moderate performance ($r = 0.747$) but with $27\%$ higher RMSE than SpiRadar.

The FEV$_1$/FVC ratio proved challenging for all methods, which is expected given that errors in both underlying parameters compound in the ratio. LinReg and SpiRadar achieved equal correlation ($r = 0.613$), but SpiRadar demonstrated the overall lowest error with the  MAE ($8.897\%$) and RMSE ($10.918\%$). The Direct method's poor performance ($r = 0.375$) confirms that accurate ratio estimation requires proper curve reconstruction rather than direct parameter regression.

PEF estimation, which is inherently variable since flow represents the derivative of volume, distinguished SpiRadar's capabilities with $r = 0.661$, and RMSE of 0.902 [L], substantially exceeding other methods ($r$ of 0.394-0.549 and RMSE of 1.016-1.245). 

\begin{figure}[htbp!]  
\begin{center}
\subfigure{\includegraphics[width=0.48\textwidth]{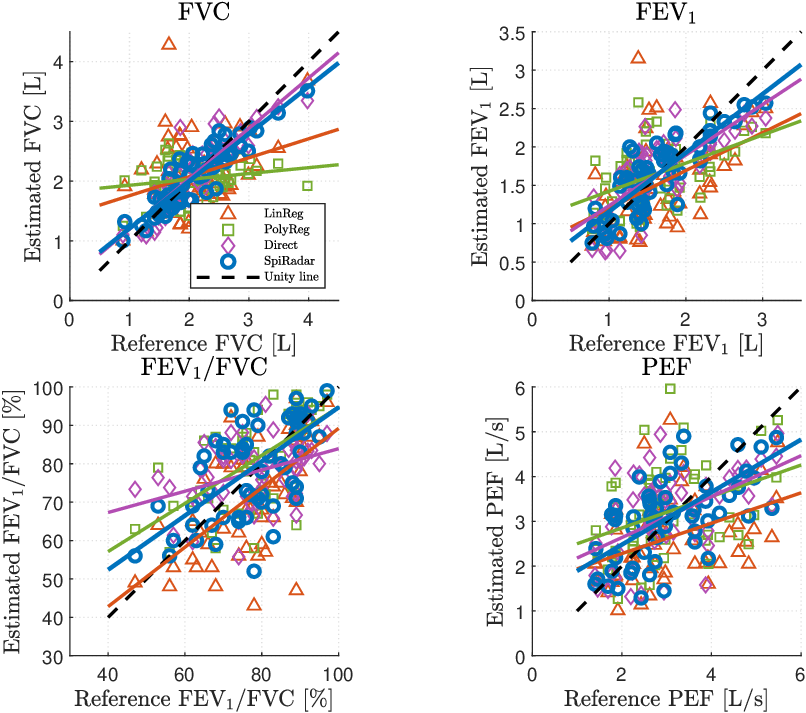}}\vspace{-0.2cm}
\end{center} 
\vspace{-0.2cm}
\caption{Correlation analysis between reference and estimated spirometry parameters by all compared methods.}   
\label{fig:correlation}
\vspace{-0.4cm}
\end{figure}

Fig.~\ref{fig:correlation} presents scatter plots illustrating the correlation analysis. SpiRadar’s estimates (in blue) clustered closely along the unity line for both FVC and FEV$_1$, with regression slopes approaching 1.0. In contrast, LinReg and PolyReg exhibited systematic underestimation for FVC, FEV$_1$, and PEF, although this bias was partially compensated in the FEV$_1$/FVC ratio. The Direct method demonstrated moderate performance for FVC, FEV$_1$, and PEF; however, its FEV$_1$/FVC estimates showed minimal correlation. Overall, SpiRadar achieved the highest $r$ correlation across all parameters.

\begin{figure}[htbp!]  
\begin{center}
\subfigure{\includegraphics[width=0.48\textwidth]{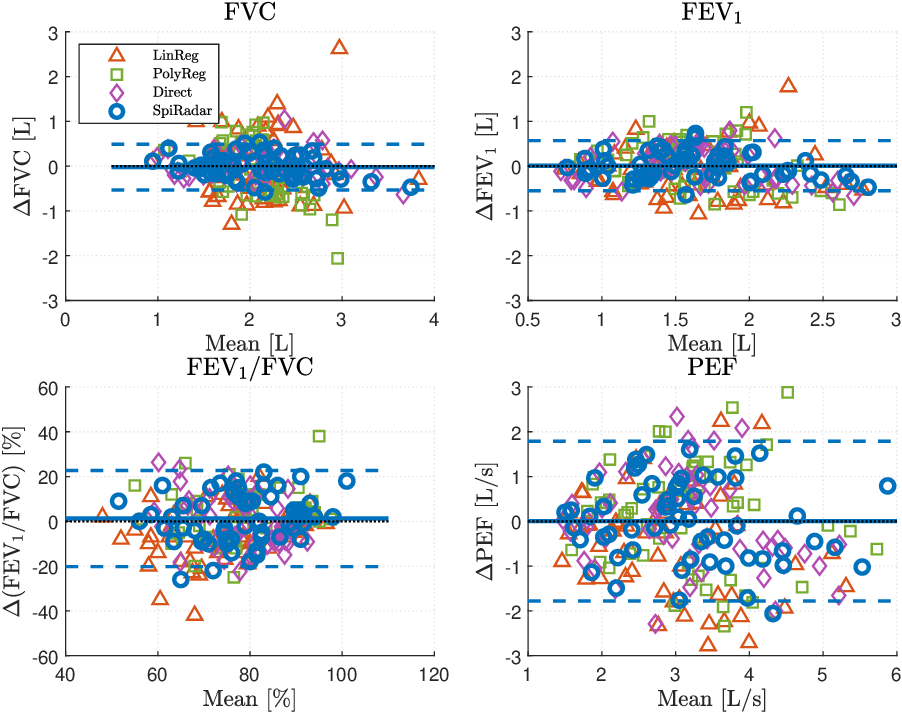}}\vspace{-0.2cm}
\end{center} 
\vspace{-0.2cm}
\caption{Bland-Altman analysis between reference and estimated spirometry parameters by all compared methods.}   
\label{fig:bland_altman}
\vspace{-0.4cm}
\end{figure}

Fig.~\ref{fig:bland_altman} presents Bland-Altman plots of the compared methods. SpiRadar (in blue) demonstrated the tightest clustering around zero bias for FVC and FEV$_1$, with 95\% limits of agreement within ±0.5 [L]. The FEV$_1$/FVC ratio showed proportional errors for all methods, with SpiRadar maintaining the narrowest limits of agreement (±20\%). The challenging PEF estimation revealed SpiRadar's consistent performance across the full range (1-6 [L/s]), while other methods exhibited increasing biases at flow rates $>3$ [L/s].

\begin{figure}[htbp!]  
\begin{center}
\subfigure{\includegraphics[width=0.48\textwidth]{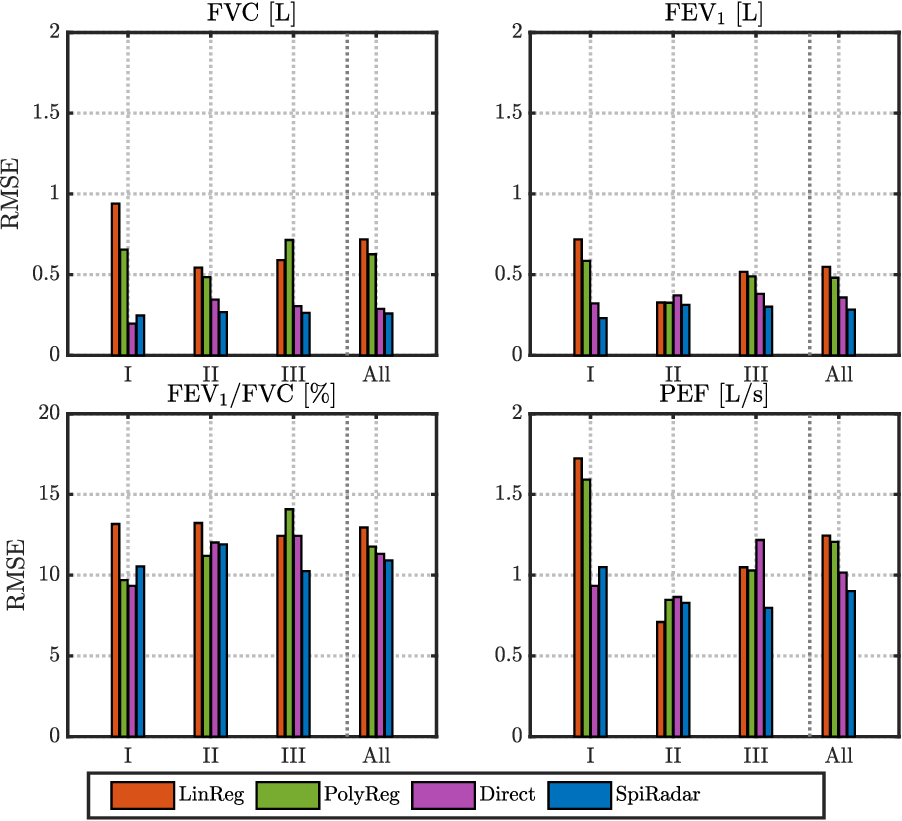}}\vspace{-0.2cm}
\end{center} 
\vspace{-0.2cm}
\caption{RMSE performance across clinical groups for spirometry parameter estimation by all compared methods. Groups: I-Healthy (n=20), II-Asthma Pre-BD (n=19), III-Asthma Post-BD (n=19), All-Combined (n=58).}   
\label{fig:rmse_groups}
\vspace{-0.4cm}
\end{figure}

Fig.~\ref{fig:rmse_groups} illustrates RMSE patterns across clinical groups. SpiRadar consistently maintained the lowest overall RMSE for every parameter, with particularly low RMSE for FVC and FEV$_1$ (0.2-0.3 [L]) across all groups. LinReg and PolyReg methods showed high variability, performing worst in healthy subjects where RMSE approached 1 [L], likely due to the greater dynamic range of patterns in normal spirometry. For FEV$_1$/FVC ratio and PEF, SpiRadar achieved the lowest overall RMSE and reached its best performance in Group III (post-BD). This highlights SpiRadar’s robustness in capturing flow dynamics in asthmatic subjects. 




\begin{table}[htbp]
\centering
\caption{Performance Metrics for Spirometry Parameter Estimation}
\label{tab:performance_metrics}
\small
\begin{tabular}{lccccl}
\toprule
\multirow{2}{*}{Method} & \multirow{2}{*}{Metric} & \multicolumn{4}{c}{Parameter} \\
\cmidrule{3-6}
 & & FVC & FEV$_1$ & FEV$_1$/FVC & PEF \\
\midrule
\multirow{3}{*}{LinReg}
 & $r$ & 0.306 & 0.489 & \textbf{0.613} & 0.413 \\
 & MAE & 0.576 & 0.436 & 10.034 & 0.969 \\
 & RMSE & 0.718 & 0.548 & 12.952 & 1.245 \\[3pt]
\multirow{3}{*}{PolyReg}
 & $r$ & 0.184 & 0.488 & 0.559 & 0.394 \\
 & MAE & 0.515 & 0.395 & 9.086 & 1.010 \\
 & RMSE & 0.626 & 0.481 & 11.767 & 1.206 \\[3pt]
\multirow{3}{*}{Direct}
 & $r$ & 0.881 & 0.747 & 0.375 & 0.549 \\
 & MAE & 0.221 & 0.302 & 9.644 & 0.828 \\
 & RMSE & 0.288 & 0.358 & 11.316 & 1.016 \\[3pt]
\multirow{3}{*}{SpiRadar}
 & $r$ & \textbf{0.902} & \textbf{0.846} & \textbf{0.613} & \textbf{0.661} \\
 & MAE & \textbf{0.213} & \textbf{0.237} & \textbf{8.897} & \textbf{0.754} \\
 & RMSE & \textbf{0.259} & \textbf{0.283} & \textbf{10.918} & \textbf{0.902} \\
\bottomrule
\end{tabular}
\\\vspace{1ex}
\footnotesize{$r$: Pearson correlation coefficient; MAE: Mean Absolute Error; RMSE: Root Mean Square Error. [L] for FVC and FEV$_1$, [\%] for FEV$_1$/FVC, [L$/s$] for PEF; Best values in \textbf{bold}.}
\end{table}

\subsection{Bronchodilator Treatment Response}
\begin{figure}[htbp!]  
\begin{center}
\subfigure{\includegraphics[width=0.40\textwidth]{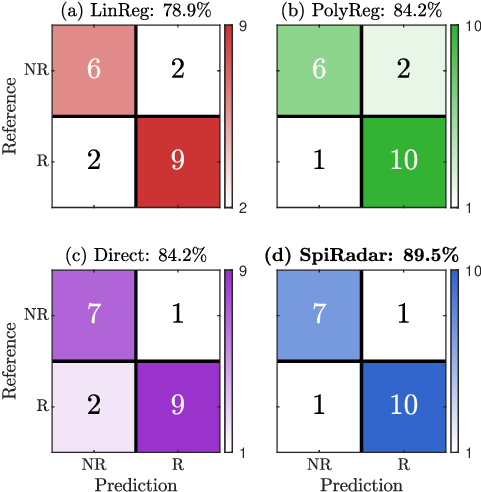}}\vspace{-0.2cm}
\end{center} 
\vspace{-0.2cm}
\caption{Confusion matrices of the compared methods for BDR classification. Each matrix shows true (by spirometry reference) versus predicted (by radar) classifications. R: responders, NR: non-responders}   
\label{fig:confusion_matrices}
\vspace{-0.4cm}
\end{figure}

As detailed in the introduction, accurate classification of BDR is essential for asthma diagnosis and management \cite{waalkens1993assessment,graham2019standardization}. Fig.~\ref{fig:confusion_matrices} presents confusion matrices for the compared methods in classifying treatment responders, defined as asthmatic patients achieving $\geq12\%$ FEV$_1$ improvement following BD administration (post-BD trial) relative to their pre-BD trial baseline  \cite{jat2013spirometry}. The analysis evaluated 19 paired pre-post measurements from asthma patients.

SpiRadar achieved the highest classification accuracy of 89.5\%, correctly identifying 17 of 19 cases with 10 true positives and 7 true negatives, and only 1 false positive and 1 false negative. LinReg achieved 78.9\% accuracy with 2 false negatives and 2 false positives, while PolyReg and Direct methods both achieved 84.2\% accuracy with contrasting error patterns.

Table \ref{tab:treatment_response} quantifies the classification performance metrics. SpiRadar achieved optimal balance with 90.9\% sensitivity and 87.5\% specificity, both critical for clinical decision-making. The lower sensitivity of LinReg and Direct (81.8\%) indicates an increased risk of missing true responders, while PolyReg matched SpiRadar's sensitivity but with reduced specificity (75.0\%), potentially leading to overtreatment. The Direct method achieved high specificity (87.5\%) but at the cost of sensitivity, missing 2 of 11 responders. 

SpiRadar's ability to accurately detect the 12\% improvement threshold despite radar measurement challenges, confirms its potential for clinical deployment in asthma diagnosis and monitoring protocols. 



\begin{table}[htbp]
\centering
\caption{Treatment Response Classification Performance}
\label{tab:treatment_response}
\begin{tabular}{lccc}
\toprule
Method & Sensitivity & Specificity & Accuracy \\
 & (\%) & (\%) & (\%) \\
\midrule
LinReg & 81.8 & 75.0 & 78.9 \\
PolyReg & \textbf{90.9} & 75.0 & 84.2 \\
Direct & 81.8 & \textbf{87.5} & 84.2 \\
\textbf{SpiRadar} & \textbf{90.9} & \textbf{87.5} & \textbf{89.5} \\
\bottomrule
\end{tabular}
\\\vspace{1ex}
\footnotesize{Treatment response is defined as $\geq$12\% FEV$_1$ post-BD improvement from pre-BD baseline \cite{jat2013spirometry}. Best values in \textbf{bold}.}
\end{table}

\subsection{Performance Sensitivity and Parameter Tuning}
We analyze SpiRadar's sensitivity to key design parameters and validate the effectiveness of preprocessing components on the estimation performance.
\subsubsection{Polynomial Order Analysis}
\label{poly_order_analysis}
\begin{figure}[htbp!]  
\begin{center}
\subfigure{\includegraphics[width=0.48\textwidth]{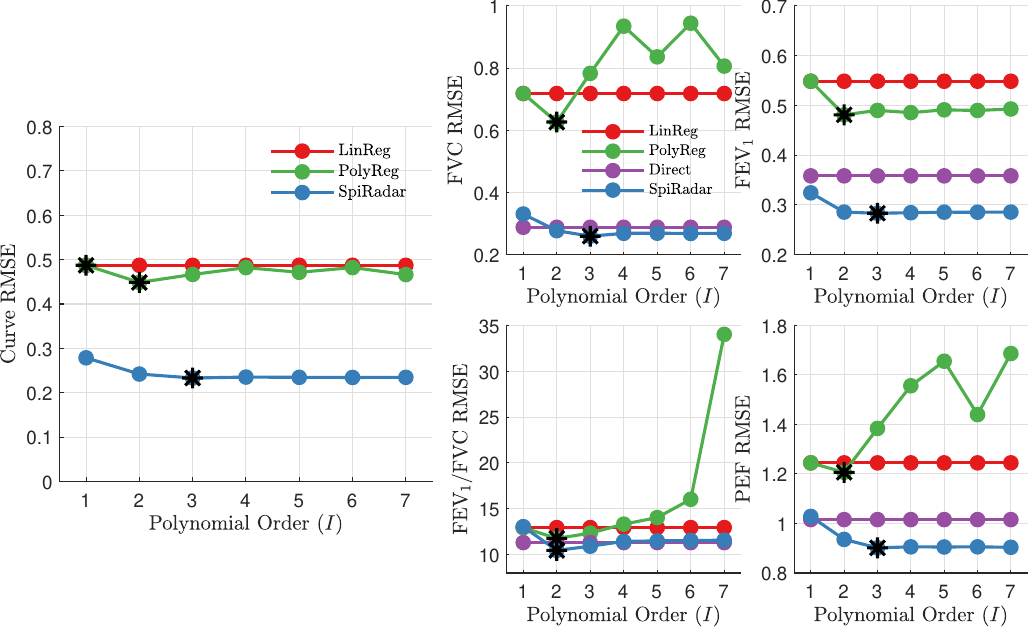}}\vspace{-0.2cm}
\end{center} 
\vspace{-0.2cm}
\caption{RMSE performance versus polynomial order (SpiRadar and PolyReg) for curve fitting and spirometry parameter estimation compared to other non-contact approaches.}   
\label{fig:RMSE_Poly}
\vspace{-0.4cm}
\end{figure}

For polynomial-based methods (SpiRadar and PolyReg), the selection of optimal polynomial order $I$ is fundamental to achieving an appropriate balance between model expressiveness and generalization capability. Fig.~\ref{fig:RMSE_Poly} illustrates the relationship between polynomial order ($I = 1, ..., 7$) and RMSE performance for both curve fitting and parameter estimation tasks, with LinReg and Direct methods included as baselines for comparison, as they are not affected by polynomial order.

The analysis revealed distinct optimization characteristics between methods. PolyReg achieved its optimal performance, indicated by an asterisk at $I = 2$, demonstrating marginal improvement over LinReg while maintaining reasonable generalization. Conversely, SpiRadar exhibited superior performance across all polynomial orders $\geq2$, achieving its minimum curve fitting RMSE of 0.23 at $I = 3$ (18\% reduction relative to the linear case $I = 1$). Notably, SpiRadar's performance plateaued beyond $I = 3$, suggesting that third-order polynomials provide sufficient representational capacity for spirometry dynamics without introducing overfitting artifacts.

Parameter-specific analysis revealed differential sensitivities to polynomial order, with PolyReg exhibiting significantly greater instability than SpiRadar. For FVC, FEV$_1$/FVC ratio, and PEF estimations, PolyReg demonstrated increasing performance degradation for $I \geq 4$, characterized by erratic RMSE fluctuations indicative of overfitting. This instability was particularly pronounced in FEV$_1$/FVC ratio estimation, where PolyReg's RMSE varied dramatically between 12\% and 35\%. In PEF estimation, PolyReg's RMSE oscillated between 1.2-1.7 L/s across polynomial orders, while SpiRadar maintained a stable 0.9 L/s for $I \geq 3$

SpiRadar maintained stable performance across higher orders, consistently outperforming alternative methods with optimal results at $I = 3$ for most parameters. This performance advantage underscores the robustness of the feature-dependent polynomial approach employed by SpiRadar. Based on these empirical findings, we selected $I = 3$ for SpiRadar and $I = 2$ for PolyReg for evaluating the methods' performance.

\subsubsection{LASSO Regularization Analysis}
\begin{figure}[htbp!]  
\begin{center}
\subfigure{\includegraphics[width=0.48\textwidth]{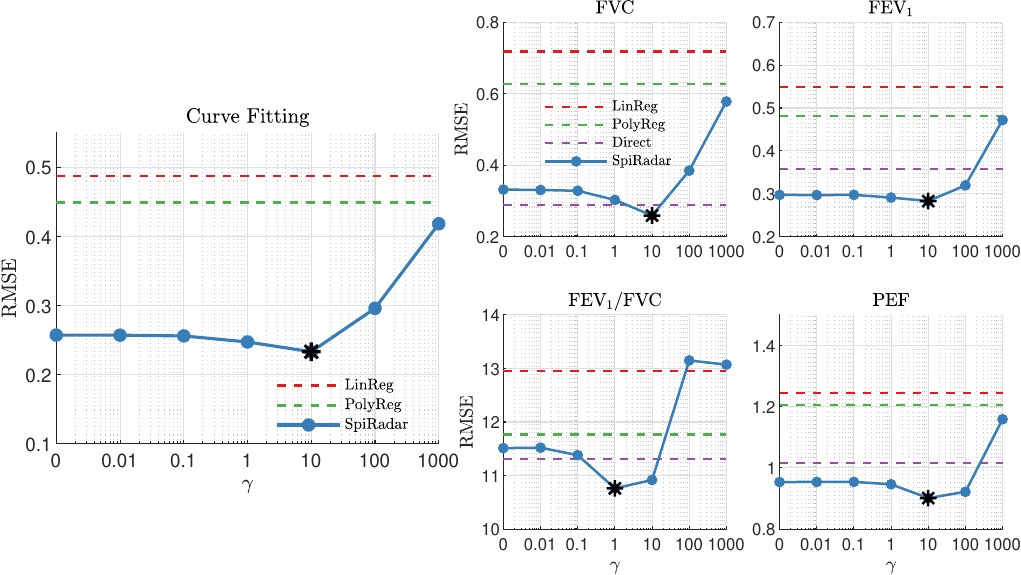}}\vspace{-0.2cm}
\end{center} 
\vspace{-0.2cm}
\caption{RMSE performance versus SpiRadar's LASSO regularization parameter $\gamma$ (\ref{LASSO}) compared to other non-contact approaches.}
\label{fig:FISTA_reg_analysis}
\vspace{-0.4cm}
\end{figure}

The sparse representation framework requires selecting the regularization parameter $\gamma$ (\ref{LASSO}) to balance sparsity with reconstruction quality. Fig. \ref{fig:FISTA_reg_analysis} illustrates SpiRadar's performance across $\gamma$ values from 0 to 1000, with competing techniques shown as horizontal baselines.

The analysis revealed $\gamma = 10$ as optimal for most parameters, achieving curve fitting RMSE of 0.23 L (vs. LinReg: 0.49 L, PolyReg: 0.44 L), FEV$_1$ RMSE of 0.29 L (22\% improvement over the best baseline, Direct: 0.37 L), and PEF RMSE of 0.91 L/s (15\% improvement over Direct: 1.05 L/s). For the FEV$_1$/FVC ratio, the optimal value was $\gamma = 1$ (RMSE: 10.7\%), though performance at $\gamma = 10$ (RMSE: 10.9\%) was negligible different.

SpiRadar maintained stable performance for $\gamma$ values up to 10 across all metrics, demonstrating robust parameter selection. Significant performance degradation occurred at $\gamma = 1000$ for all parameters, indicating over-regularization. The consistent superiority over baseline methods across the stable regularization range validates the sparse representation approach. Based on this analysis, we selected $\gamma = 10$ for all SpiRadar evaluations.

\subsubsection{Feature Vector Selection}
\label{feature_analysis}
The feature vector $\boldsymbol{\alpha}$ enables subject-specific adaptation through the feature-dependent polynomial model (\ref{eq:poly_model}-\ref{eq:beta_features}). The initial formulation included radar-derived FVC$^{\text{R}}$ (calculated as $\max_l\{v_{ex}[l]\}$) alongside the features in (\ref{final_feature_vector}).

Statistical analysis using ANOVA F-statistics and $r$ coreelation revealed that FVC$^{\text{R}}$ showed limited predictive utility, demonstrating moderate correlation only with the target FVC parameter ($r=0.329$) while having weak correlations with FEV$_1$, ratio, and PEF parameters ($r<0.2$), which risks overfitting and reduce model robustness. In contrast, other radar-derived features showed broader utility: FEV$_1$$^{\text{R}}$ demonstrated moderate correlations across multiple parameters ($r=0.31$–$0.45$), and ratio$^{\text{R}}$ was particularly valuable for predicting the FEV$_1$/FVC ratio ($r=0.592$). Anthropometric features showed strong predictive relationships, with age achieving correlations of $r=0.832$ (FVC), $r=0.686$ (FEV$_1$), and $r=0.570$ (PEF).

Removing FVC$^{\text{R}}$ improved overall performance: for example, curve fitting RMSE decreased from 0.247 to 0.233 L, while the correlation of the parameters FVC, FEV$_1$ and PEF improved from 0.881 to 0.902, from 0.837 to 0.846, and from 0.640 to 0.661, respectively. The optimized feature set eliminated the narrow predictor while maintaining physiological relevance and enhancing model generalization.

\subsubsection{Interpolation Correction Effect}
\label{sec:signal_correction}
As described in the preprocessing pipeline (Section \ref{Pre_Proc}), radar-based thoracic signals may exhibit artifacts during spirometry maneuvers. Here we analyze the effect of our logarithmic interpolation correction designed to follow the physiological nature of expiratory behavior.

\begin{figure}[htbp!]  
\begin{center}
\subfigure{\includegraphics[width=0.48\textwidth]{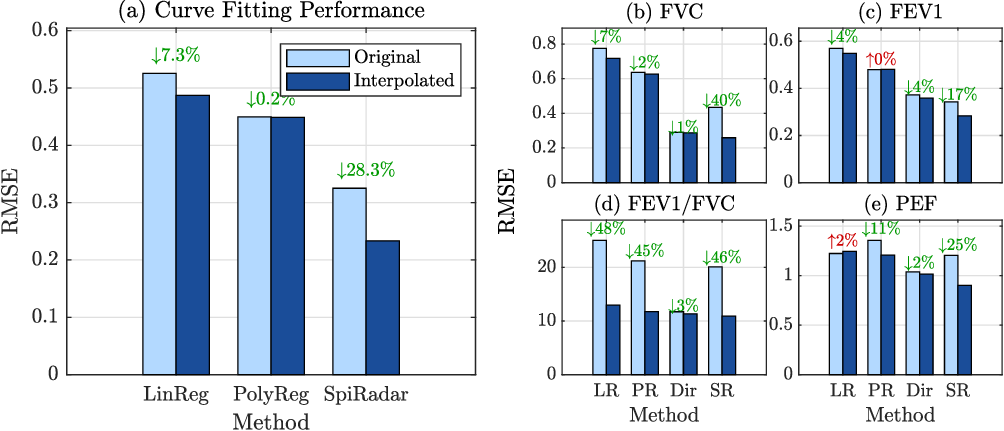}}\vspace{-0.2cm}
\end{center} 
\vspace{-0.2cm}
\caption{RMSE improvement percentages after logarithmic interpolation correction for curve fitting and spirometry parameter estimation (FVC, FEV$_1$, FEV$_1$/FVC, PEF) comparing SpiRadar to other non-contact approaches. LR: LinReg, PR: PolyReg, Dir: Direct, SR: SpiRadar.}
\label{fig:interpolation_improvement_analysis_updated}
\vspace{-0.4cm}
\end{figure}

Comprehensive analysis across all 58 trials revealed consistent benefits of interpolation (Fig.~\ref{fig:interpolation_improvement_analysis_updated} and Table \ref{tab:interpolation_improvement_summary}). As seen in Table \ref{tab:interpolation_improvement_summary}, SpiRadar's curve fitting RMSE improved by 28.3\%, while spirometry parameter estimation showed even greater enhancements. FVC estimation benefited most significantly with 40.4\% RMSE reduction and 23.4\% correlation improvement. The challenging FEV$_1$/FVC ratio showed the largest relative improvements with 45.7\% RMSE reduction and 58.6\% correlation increase.

In comparison to the other techniques, Fig. \ref{fig:interpolation_improvement_analysis_updated} shows that SpiRadar consistently achieved the largest RMSE improvements across most spirometry parameters, suggesting that the sparse optimization framework effectively exploits improved signal quality. While other methods showed more modest gains, interpolation was beneficial across most parameters, with very small to no degradation in the remaining. The overall benefits validate interpolation as a valuable preprocessing step that could enhance performance for various radar-based spirometry techniques.

\begin{figure}[htbp!]  
\begin{center}
\subfigure{\includegraphics[width=0.48\textwidth]{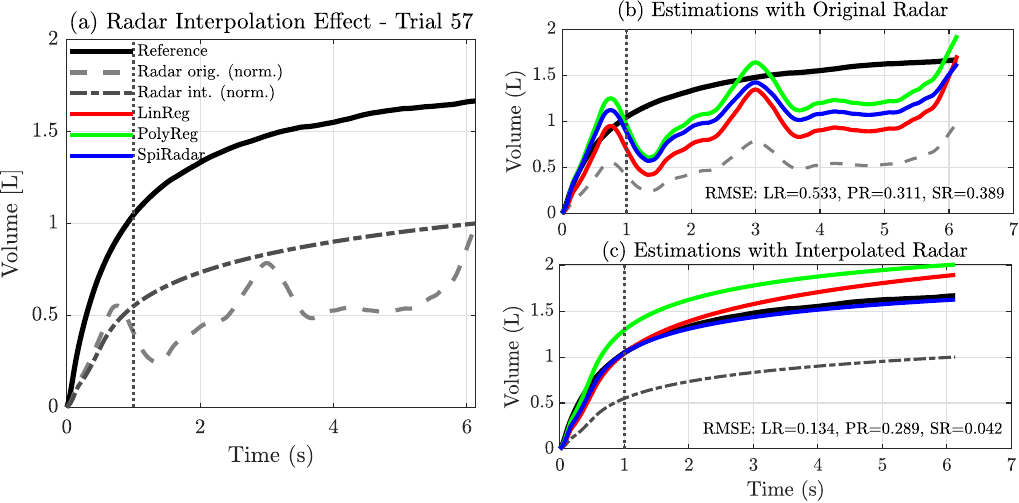}}\vspace{-0.2cm}
\end{center} 
\vspace{-0.2cm}
\caption{Interpolation effect demonstration for Trial 57 (Asthma Pre-BD). (a) Normalized original radar signal with artifacts before interpolation correction (light gray dashed) and after (dark gray dashed), compared to reference spirometry (black solid). (b,c) Volume curve estimations using original (b) versus corrected (c).}
\label{fig:interpolation_effect_trial_57}
\vspace{-0.4cm}
\end{figure}

Fig.~\ref{fig:interpolation_effect_trial_57} demonstrates the interpolation effect on Trial 57 (Asthma Pre-BD). Panel (a) shows the original radar signal (light gray dashed line) exhibiting significant oscillations and artifacts that deviate from the smooth physiological pattern of the reference spirometry (black solid line). In contrast, the interpolated radar signal (Dark gray dashed line) exhibits improved monotonic curve shape. Panels (b) and (c) compare volume curve estimations using original versus interpolated radar signals, respectively. With interpolation, all methods show dramatically improved alignment with the reference curve, with SpiRadar (blue) achieving the best performance. The RMSE improvements were substantial: LinReg decreased from 0.533 to 0.134 L (75\% reduction), PolyReg from 0.311 to 0.289 L (7\% reduction), and SpiRadar from 0.389 to 0.042 L (89\% reduction and smallest RMSE).

\begin{table}[htbp]
\centering
\caption{SpiRadar Improvement Due Interpolation}
\label{tab:interpolation_improvement_summary}
\begin{tabular}{l|ccc}
\hline
Parameter & $r$ (\%) & MAE (\%) & RMSE (\%) \\
\hline
Curve Fitting & -- & -- & 28.3 \\
\hline
FVC & 23.4 & 30.7 & 40.4 \\
FEV$_1$ & 5.9 & 13.7 & 17.3 \\
FEV$_1$/FVC & 58.6 & 35.1 & 45.7 \\
PEF & 9.5 & 26.3 & 25.2 \\
\hline
\end{tabular}
\end{table}

\section{Discussion}
\label{sec:discussion}
This study presents the first comprehensive evaluation of mmWave radar-based spirometry in a pediatric cohort, demonstrating that SpiRadar achieves clinically relevant accuracy for both curve reconstruction and parameter estimation. The feature-dependent polynomial model effectively captures subject-specific variations while maintaining generalizability across the pediatric age range. The $>49\%$ RMSE reduction in curve fitting compared to other curve regression approaches, combined with parameter $r$ correlations exceeding 0.84 for key spirometric indices, represents a significant advancement in non-contact respiratory monitoring.

The parameter estimation results demonstrate particular strength in FVC ($r = 0.902$, RMSE = 0.259 L) and FEV$_1$ ($r = 0.846$, RMSE = 0.283 L). Notably, SpiRadar maintained robust performance even in challenging clinical scenarios, including asthma measurements where obstructed flow patterns typically complicate analysis. The 89.5\% accuracy in BDR classification, with balanced sensitivity (90.9\%) and specificity (87.5\%), indicates potential for clinical deployment in asthma management protocols.

The optimal polynomial order of $I = 3$ provides an advantageous balance between model expressiveness and generalization capability, avoiding the overfitting artifacts observed in higher-order polynomial approaches. The sparse optimization framework demonstrated robust performance across three orders of regularization magnitude, indicating practical parameter tuning requirements. Feature selection analysis revealed that removing the radar-derived FVC$^{\text{R}}$ for the feature vector  $\boldsymbol{\alpha}$ (\ref{final_feature_vector}) improved overall performance, as this feature showed limited cross-parameter utility compared to other features. The logarithmic interpolation correction addressed physiological constraints inherent in expiratory maneuvers, yielding improvements ranging from 17.3\% (FEV$_1$) to 45.7\% (FEV$_1$/FVC ratio) of RMSE reduction across spirometric parameters.

Unlike previous studies that focused primarily on tidal breathing or required subject-specific calibration, SpiRadar achieves generalization through LOOCV without individual calibration. The superior performance compared to direct parameter estimation (Direct) validates two critical advantages: first, curve reconstruction preserves essential diagnostic morphological information needed for distinguishing obstructive from restrictive patterns and assessing disease severity; second, the intermediate curve reconstruction step actually improves parameter estimation accuracy compared to the direct regression approach, demonstrating that the physiologically-motivated modeling pathway enhances overall system performance.

Several limitations warrant consideration. The dataset, while representing the first comprehensive pediatric radar spirometry evaluation, encompasses only 39 subjects from a single medical center. The study population was limited to children aged 6-18 years, and generalization to adult populations or respiratory conditions other than asthma requires further validation. Additionally, the comparison with traditional spirometry was conducted under controlled conditions, and real-world deployment may introduce additional challenges.

\section{Conclusion}
\label{sec:conclusion}
We presented SpiRadar, a comprehensive mmWave radar framework for non-contact spirometry that achieved clinical-grade performance on a pediatric cohort. Through rigorous subject-level LOOCV, the framework demonstrated curve reconstruction RMSE of 0.23 L, strong correlations for key spirometric indices (FVC: $r=0.902$, FEV$_1$: $r=0.846$), and accurate treatment response classification (89.5\% accuracy) without requiring individual calibration. The integrated methodological approach combines physiologically-motivated preprocessing, feature-dependent polynomial transformation, and sparse optimization to enable generalization across diverse subjects while preserving diagnostic curve morphology essential for clinical interpretation. These results establish radar-based spirometry as a viable non-contact alternative for pulmonary function assessment, with particular promise for pediatric asthma monitoring and home-based patient care.

\bibliographystyle{IEEEtran}
\bibliography{References}

\end{document}